\documentclass[preprints,article,accept,moreauthors,pdftex]{Definitions/mdpi} 

\firstpage{1} 
\pubvolume{1}
\issuenum{1}
\articlenumber{0}
\pubyear{2021}
\copyrightyear{2021}
\datereceived{} 
\dateaccepted{} 
\datepublished{} 
\hreflink{https://doi.org/} 
\pdfoutput=1

\usepackage{dsfont}
\usepackage{physics}

\usepackage{todonotes}
\renewcommand{\vec}[1]{\mathbf{#1}}
\newcommand{\mat}[1]{\mathbf{#1}}
\newcommand{\dt}{\delta\tau}

\Title{Polaron-Depleton Transition in the Yrast Excitations of a One-Dimensional Bose Gas with a Mobile Impurity}

\TitleCitation{Polaron-Depleton Transition in Yrast Excitations of One-Dimensional Bose Gases with a Mobile Impurity}

\Author{Mingrui Yang $^{1,3,}$*\orcidA{}, Matija \v{C}ufar$^{1}$\orcidB{}, Elke Pahl $^{2,3}$\orcidC{} and Joachim Brand $^{1}$\orcidD{}}

\AuthorNames{Mingrui Yang, Matija \v{C}ufar, Elke Pahl and Joachim Brand}

\AuthorCitation{Yang, M.; \v{C}ufar, M.;  Pahl, E.; Brand, J.}

\address{%
$^{1}$ \quad Dodd-Walls Centre for Photonic and Quantum Technologies, New Zealand Institute for Advanced Study and Centre for Theoretical Chemistry and Physics, Massey University, Auckland 0632, New Zealand\\
$^{2}$ \quad Department of Physics, University of Auckland, Auckland 1010, New Zealand\\
$^{3}$ \quad MacDiarmid Institute for Advanced Materials and Nanotechnology, Wellington 6140, New Zealand
}

\corres{Correspondence: M.Yang4@massey.ac.nz}

\abstract{
We present exact numerical data for the lowest-energy momentum eigenstates (yrast states) of a {repulsive} spin impurity in a one-dimensional Bose gas using full configuration interaction quantum Monte Carlo (FCIQMC). As a stochastic extension to exact diagonalization it is well suited for the study of yrast states of a lattice-renormalized model for a quantum gas.
Yrast states carry valuable information about the dynamic properties of slow-moving mobile impurities immersed in a many-body system.
Based on the energies and the first and second order correlation functions of yrast states, we identify different dynamical regimes and the transitions between them:
The polaron regime, where the impurity's motion is affected by the Bose gas through a renormalized effective mass; a regime of a gray soliton that is weakly correlated with a stationary impurity, and the depleton regime, where the impurity occupies a dark or gray soliton. Extracting the depleton effective mass reveals a super heavy regime where the magnitude of the (negative) depleton mass exceeds the mass of the finite Bose gas. 
}


\begin{document}




\section{Introduction}

The study of a single quantum impurity in a surrounding many-body medium has fascinated scientists for many decades
\cite{Emin2012,Alexandrov2010}. Beyond the historical interest around the influence of the crystal lattice on the motion of  an electron -- the original ``polaron'' \cite{Landau1948a}, or impurity atoms in superfluid helium \cite{Bardeen1966}, there has recently been a surge of interest in the field of ultracold atoms, where interactions can be readily tuned with the help of Feshbach resonances \cite{Chin2010} and excitation spectra probed with spectroscopic methods \cite{Vale2021a}.
A particular focus of experimental scrutiny has been the Bose polaron, where an impurity atom is coupled with a bosonic bath \cite{Jorgensen2016,Hu2016,Yan2020,Skou2021}.

Restricting the dimensionality to one spatial dimension provides access to the special physics of one dimensional quantum liquids \cite{Imambekov2012,Cherny2012}, where impurities have been predicted to undergo Bloch oscillations \cite{Gangardt2009,Schecter2012}: Due to the periodicity of the dispersion relation, an impurity  experiencing a weak force  periodically alters its excitation state without contributing to transport in real space, as originally predicted \cite{Bloch1929} and later observed  \cite{Feldmann1992,BenDahan1996} for a particle in an external lattice potential.  The prediction of Bloch oscillations in a one-dimensional quantum liquid even in the absence of a periodic potential \cite{Gangardt2009,Schecter2012} was debated \cite{Gamayun2014,Schecter2016}, but eventually confirmed in an experiment with spin impurities in a one-dimensional gas of cesium atoms \cite{Meinert2017}. Other experiments probing impurity physics in one-dimensional quantum gases also employed spin impurities (where the impurity atoms have the same mass and only differ in a spin quantum number) \cite{palzer09,Fukuhara2013a}, or different types of atoms \cite{Catani2012,Spethmann2012}.

While experimental studies have started, the quantitative understanding of impurity physics in a one-dimensional Bose gas is far from complete.
The different theoretical approaches to the problem range 
from mean-field theory \cite{Kain2018,Panochko2019} and related variational theory \cite{Dutta2013,Koutentakis2021,Seetharam} via the path-integral approach \cite{Ichmoukhamedov2019, Jager2021a}, the renormalization group  \cite{Volosniev2017,Grusdt2015,Isaule2021a} and flow equation \cite{Brauneis2021} method to multiconfiguration time-dependent Hartree  \cite{Mistakidis2019} and quantum Monte Carlo methods \cite{Grusdt2017,Ardila2015,PenaArdila2019a,Parisi2017}.

Many of the mentioned works focus on ground state properties  and effective mass of the one-dimensional Bose polaron  with either zero or very small total momentum. Although the dynamics of impurities has also been actively studied \cite{Mistakidis2019,Koutentakis2021,Seetharam,Schmidt2021a}, limited understanding has been achieved on the full dispersion relation of a Bose gas coupled with a mobile impurity. There are analytical results on the dispersion relations restricted to specific models, such as the Yang-Gaudin model \cite{Ristivojevic2021a}, and the Luttinger liquid \cite{Lamacraft2009}. 

In a homogeneous one-dimensional gas, e.g.~in a ring geometry with periodic boundaries,
translational invariance makes the total momentum a good quantum number. This allows for the study of yrast states, which are eigenstates with lowest energy at given momentum.
Yrast states are stable as long as momentum is conserved, while adiabatic passage through the yrast states of different momentum is responsible for the Bloch oscillation phenomena of Refs.~\cite{Gangardt2009,Meinert2017}.
The yrast states of a bosonic superfluid in the absence of impurities are intimately connected \cite{Kulish1976,Kanamoto2008,Kanamoto2010,Jackson2011,Fialko2012,Sato2012,Syrwid2015,Shamailov2019} to localized nonlinear waves known as dark solitons \cite{Tsuzuki1971}. Dark solitons are ubiquitous features of superfluids, which can be characterized by a localized density depression and a phase jump \cite{Shamailov2016,Shamailov2019,Syrwid2021a}.

When a {repulsive} impurity is introduced into the Bose gas, two different low-energy configurations can {exist} depending on the momentum: At a lower momentum, the impurity moves relative to the quantum gas forming a polaron. At  higher momentum, the bulk of the momentum is taken by the Bose gas forming a gray or dark soliton modified by the presence of the impurity. This situation was named the ``depleton'' in Refs.~\cite{Schecter2016,Schecter2012}. 

For strongly correlated impurity problems that are outside the reach of analytically solvable models, quantum  Monte Carlo (QMC) methods have proven invaluable tools   \cite{Astrakharchik2013,Grusdt2017,Ardila2015,PenaArdila2019a,Parisi2017}.
In this work we employ the full configuration interaction quantum Monte Carlo (FCIQMC) \cite{Booth2009,Cleland2010} method. It can be seen as  a natural stochastic extension to the exact diagonalization method, which allows one to treat a larger Hilbert space that could otherwise not fit into the computer memory. While different in detail it is similar in sprit to earlier versions of projector Monte Carlo methods \cite{Kalos2007} in sampling the ground state wave function.
When applied to a translationally invariant Hamiltonian in momentum space, FCIQMC has the advantage over other QMC methods like diffusion Monte Carlo (which is formulated in real space) or auxiliary field QMC that momentum is strictly conserved in each elementary stochastic operation. Thus yrast states can be obtained easily by projection onto the lowest-energy state within a total-momentum sector  starting from an initial state with the same momentum.
Moreover, FCIQMC mitigates the sign problem by walker annihilation in many systems when a sufficient number of walkers is present \cite{Spencer2012}. Additionally the initiator approximation \cite{Cleland2010} can be applied to suppress the sign problem with the trade-off that a small initiator bias is introduced. 

The FCIQMC method was originally developed for fermionic many-body problems. It has been applied to  the electronic structure of molecules and solids \cite{Booth2011,Cleland2012a,Booth2013} and the Hubbard model \cite{Schwarz2015,Yun2017,Yun2021}.
Recently it was used to study the yrast states in a superfluid of spin-$\frac{1}{2}$ fermions \cite{Ebling2021a}.
Here we use FCIQMC for the first time to quantitatively study the physics of a bosonic many-body problem,
while previously bosonic Hamiltonians were employed when developing and analysing the FCIQMC procedures \cite{Yang2020, Brand2021}.

In this work we use the FCIQMC method to obtain numerical results for the yrast states of  a one-dimensional Bose gas containing a {repulsive} spin impurity.
%
%
%
We characterize the polaron and depleton regimes of the yrast dispersion, as well as the transitions between them, by examining the energies and the first and second order correlation functions of yrast states.
The extracted depleton effective mass reveals a super-heavy regime where the magnitude of the (negative) depleton mass exceeds the mass of the finite Bose gas. The results also show that the depleton picture becomes inadequate for smaller impurity-boson interactions where the impurity and Bose-gas motion decouples.

In Sec.~\ref{sec:model} we introduce the lattice discretized model Hamiltonian with renormalized coupling constants to be studied in this work. 
The FCIQMC algorithm and modifications made for treating bosonic Fock states along with implementation details and parameter choices are discussed in Sec.~\ref{sec:method}.
Numerical results on properties of the yrast states and their interpretation in terms of the different  physical regimes are presented in Sec.~\ref{sec:results}
starting with the yrast dispersion in Sec.~\ref{sec:yrast_disp}. This is followed by the impurity momentum and  the impurity-boson correlation function in Secs.~\ref{sec:Pimp} and \ref{sec:g2}, respectively. The effective mass in the soliton/depleton region of the dispersion is reported in Sec.~\ref{sec:mass} and the spin-flip energies in Sec.~\ref{sec:spin-flip}.
Finally, we draw conclusions in Sec.~\ref{sec:conclusions} and we outline possible future prospects of this work. Appendix \ref{sec:Bias} presents data on the  elimination of systematic biases in FCIQMC, which is relevant for validating the computational method.

\section{The Model}\label{sec:model}

\subsection{The Hamiltonian in one-dimensional real space}
We consider a single impurity particle immersed in a one-dimensional interacting Bose gas of $N$ identical particles. The  Hamiltonian reads
\begin{equation}\label{eq:1d-real-ham}
    H = -\frac{\hbar^2}{2m}\sum_{i=1}^N \frac{\partial^2}{\partial x_i^2} + g_\mathrm{BB} \sum_{i<j} \delta(x_i-x_j) -\frac{\hbar^2}{2m}\frac{\partial^2}{\partial x_\mathrm{I}^2} + g_\mathrm{IB} \sum_{i=1}^N \delta(x_i-x_\mathrm{I}),
\end{equation}
where $x_i(i=1,\cdots,N)$ and $x_\mathrm{I}$ are the coordinates of the bosons and the impurity, respectively. 
We have already assumed that the impurity has the same mass $m$ as the bosons do, and will continue to do so throughout this work. This is adequate for a spin impurity where the impurity atom becomes distinguishable from the remaining bosons by changing a spin quantum number, \textit{e.g.}~changing a hyperfine quantum number of an ultracold atom.
The $N$ bosons  are interacting with a contact potential of strength $g_\mathrm{BB}$ while the interaction of the impurity particle  with the bosons is described by a contact potential of strength $g_\mathrm{IB}$. 
We consider repulsive interactions with {$g_\mathrm{BB} >0$ and $g_\mathrm{IB} \ge 0$ in order to access the physics of the polaron--depleton transition. We leave the detailed study of attractive impurities, which bind to bosons rather than to the hole-like dark soliton excitations, to future work.}

Following the definitions from previous works on polaron problems \cite{Parisi2017,Grusdt2017,Panochko2019}, we introduce the dimensionless coupling parameters
\begin{equation}\label{eq:gamma-and-eta}
    \gamma = \frac{mg_\mathrm{BB}}{\hbar^2n}, \quad \eta = \frac{mg_\mathrm{IB}}{\hbar^2n},
\end{equation}
to represent the boson--boson and boson-impurity interaction strengths, respectively. The density of the Bose gas is $n=N/L$. Note that with the impurity and the bosons having equal mass, the reduced mass, $m_r$, used in other works, becomes $m/2$.
\subsection{Lattice discretized continuum model}
For our numerical simulations we consider a finite system in a one-dimensional box of length $L$ with periodic boundary conditions. 
We discretize the model using a lattice with $M$ lattice sites with renormalized contact interactions \cite{Castin2004,Ernst2011}. In order to access yrast states numerically with FCIQMC, we use a momentum-space representation  of the Hamiltonian. In FCIQMC individual stochastic sampling steps then conserve momentum, which allows us to access the yrast states with this projector QMC method.
The spatial domain is $x \in \left(-L/2, L/2\right]$ and the lattice constant is defined as $\alpha=L/M$ for $M$ lattice points.
In this representation, the Hamiltonian reads
\begin{equation}\label{eq:2c-ham-mom}
    H^\mathrm{mom}= \sum_k \epsilon_k \hat{a}_k^\dag \hat{a}_k + 
    \sum_k \epsilon_k \hat{b}_k^\dag \hat{b}_k + 
    \frac{U}{2M} \sum_{spqr} \hat{a}_s^\dag \hat{a}_p^\dag \hat{a}_q \hat{a}_r \delta_{s+p,q+r} + 
    \frac{V}{M} \sum_{spqr} \hat{a}_s^\dag \hat{b}_p^\dag \hat{b}_q \hat{a}_r \delta_{s+p,q+r},
\end{equation}
where $\hat{a}_k^\dag (\hat{a}_k)$ are the boson creation (annihilation) operators; the corresponding operators for the impurity are $\hat{b}_k^\dag (\hat{b}_k)$. 
The plane-wave eigenstates $\langle x|\hat a^\dag_k|\mathrm{vac}\rangle = e^{-ikx/\alpha}$ 
of momentum $\hbar k\alpha$
are indexed  with the dimensionless quantum numbers 
\begin{equation}
\label{eq:ktildediscrete}
k_j =
\begin{cases}
-\pi + j \frac{2\pi}{M} 
&\quad \mbox{if $M$ is even}\\
-\pi \frac{M+1}{M} + j \frac{2\pi}{M}
&\quad \mbox{if $M$ is odd}
\end{cases}
\end{equation}
where $j \in\{1,2, \ldots, M\}$ is an integer. The kinetic energy dispersion is the same for bosons and impurity (as they have equal mass)
\begin{align}
\epsilon_k = \frac{\hbar^2 k^2}{2m \alpha^2} = \frac{1}{2}{M^2 k^2}\varepsilon_0 ,
\end{align}
where we have introduced the unit of energy that will be used throughout this work
\footnote{Note that Refs.~\cite{Parisi2017, Panochko2019} choose the Fermi energy $\varepsilon_F = \frac{\pi^2\hbar^2n^2}{2m} = \frac{\pi^2N^2}{2}\varepsilon_0$ as the energy unit. }
\begin{equation}
    \varepsilon_0 = \frac{\hbar^2}{mL^2}.
\end{equation}

The parameters $U$ and $V$ are the lattice on-site interaction strengths for boson-boson and boson-impurity, respectively.
They are renormalized to generate the correct scattering length for a two-particle scattering problem at zero energy  \cite{Ernst2011, Castin2004}:
\begin{align}
U\alpha = \frac{ g_\mathrm{BB}}{1+\frac{ g_\mathrm{BB}}{g_0}} , \quad V\alpha = \frac{ g_\mathrm{IB}}{1+\frac{ g_\mathrm{IB}}{g_0}} , 
\end{align}
where $g_0 = {\pi^2\hbar^2}/{m\alpha}$.

\subsection{Connection to mean-field theory and choice of parameters}

In the weakly interacting regime where $\gamma\ll1$, nonlinear phenomena in the Bose gas like dark and gray solitons are accurately described by the Gross-Pitaevskii equation \cite{Tsuzuki1971,Shamailov2019}. A similar mean-field treatment is also available for a Bose gas with an impurity \cite{Volosniev2017,Schecter2016}. The relevant length scale in this theory is the healing length, which is the shortest length scale on which the superfluid order parameter can change
\begin{equation}
    l_\mathrm{h} = \frac{\hbar}{\sqrt{2mg_\mathrm{BB}n}} = \frac{L}{\sqrt{2\gamma}N}.
\end{equation}
In order to obtain insights into the physics of solitons and their interaction with impurities, we need to choose the parameters of our model system such that $L\gg l_\mathrm{h}$. In order to obtain results relevant for the thermodynamic limit it would be desirable to choose both the particle number $N$ and the box size $L$ large. However, we are constrained by the fact that FCIQMC has a sign problem, which limits the maximum number of particles and modes that can be accurately computed. The sign problem also grows more severe with stronger interaction strength, which affects the off-diagonal matrix elements in the Hamiltonian of Eq.~\eqref{eq:2c-ham-mom}. As a compromise we choose to work with $N_\mathrm{tot}=20$ particles and $M=50$ modes and explore the weakly-interacting regime. This yields a ratio of box size to healing length  of $L/l_\mathrm{h} \equiv {\sqrt{2\gamma}N} \approx 12$ for $\gamma = 0.2$ and  $L/l_\mathrm{h} \approx 4$ for $\gamma=0.02$ for the two values of the Bose-gas interaction strength that we are using in this work.

\section{Computational method and simulation details}\label{sec:method}

Full configuration interaction quantum Monte Carlo is a projector quantum Monte Carlo method that can be used to determine the ground-state energies of quantum many-body systems. It was originally formulated to solve problems in quantum chemistry~\cite{Booth2009}. In this section, we describe the algorithm and some of the modifications we implemented 
to extend the algorithm to describe bosonic systems.

\subsection{Bosonic Full Configuration Quantum Monte Carlo}

In FCIQMC a basis of Fock states (occupation number basis) for $N=\sum_{i=1}^M n_i$ particles in $M$ lattice sites is used
\begin{equation}
| n_1, n_2, \dots n_M \rangle = \prod_{i=1}^{M} \frac{1}{\sqrt{n_i !}} \left (\hat{a}_i^\dagger \right )^{n_i} |\mathrm{vac} \rangle .
\label{eq:Fockstates}
\end{equation}
Within this basis the Hamiltonian is represented as a matrix $\mat{H}$ and the quantum state (many-body wave function) as a vector $\vec{c}$ containing the signed weights of the individual Fock states as coefficients.

The ground-state coefficient vector is then found in an iterative manner by repeatedly applying the  equation
\begin{equation}\label{eq:fciqmc}
  \vec{c}^{(n+1)} = \vec{c}^{(n)} + \dt\left(\mat{1}S^{(n)} - \mat{H}\right) \vec{c}^{(n)},
\end{equation}
where the parameter $\dt$ controls the size of the time step, $\vec{c}^{(n)}$ is the approximation of the eigenvector at the $n$-th time step, and $\mat{1}$ is the identity matrix. The shift $S^{(n)}$ is a real number used to keep the norm of $\vec{c}^{(n)}$ under control. It is adjusted by the following scheme:
\begin{equation}
  S^{(n+1)} = S^{(n)} -\frac{\zeta}{\dt}\ln\left(\frac{N_\mathrm{w}^{(n+1)}}{N_\mathrm{w}^{(n)}}\right)
  - \frac{\xi}{\dt}\ln\left(\frac{N_\mathrm{w}^{(n+1)}}{N_\mathrm{t}}\right),
\end{equation}
where $N_\mathrm{w}^{(n)} \equiv \Vert \vec{c}^{(n)} \Vert_1$ is the 1-norm of $\vec{c}^{(n)}$, 
$N_\mathrm{t}$ the parameter for the target norm, 
and $\xi$ and $\zeta$  parameters that control the dynamics of the shift. 
{In the steady state, the instantaneous norm $N_\mathrm{w}^{(n)}$ fluctuates around the value of $N_\mathrm{t}$~\cite{Yang2020}. It is important to control the vector norm in FCIQMC as it is a proxy for the number of (stored) non-zero elements of the coefficient vector and thus for both memory and runtime requirements of the simulation. }

Because the size of the Hilbert space grows exponentially with system size, both $\mat{H}$ and $\vec{c}$ quickly become prohibitively large. To get around this problem, we replace the matrix-vector multiplication in Eq.~\eqref{eq:fciqmc} with a stochastic sampling process. The sampling process is designed to reproduce the right hand side of Eq.~\eqref{eq:fciqmc} by expected value while at the same time replacing most coefficients in the vector with zero, such that the values do not have to be stored.
Concretely, we divide the values of the entries in $\vec{c}$ into integer units called ``walkers''. At each time step, each walker attempts to ``spawn'' to a configuration connected by a non-zero entry in the corresponding column of the matrix $\mat{H}$.

The spawning from the configuration $q$ to the configuration $r\neq q$ can be described as
\begin{equation}
  {c}_r \gets {c}_r - \frac{\dt}{p_{\mathrm{spawn}}} H_{r,q}c_q,
\end{equation}
where $\frac{1}{p_{\mathrm{spawn}}}$ is the inverse probability of picking $r$, i.e. the number of nonzero  off-diagonal entries in the $q$-th column of $\mat{H}$. If the occupation number $\vec{c}_q$ is greater than the number of non-zero entries in this column, the spawns can be performed exactly. In addition to the off-diagonal spawns, the diagonal part of the matrix-vector multiplication in Eq.~\eqref{eq:fciqmc} is performed exactly.
After a step is complete, we stochastically project the entries $v_i$ of the vector $\vec{c}$ to a threshold $t$; values $|v_i| < t$ are removed from the vector with probability $p=1 - \frac{|v_i|}{t}$. Otherwise, their value becomes $v_i=t$. In practice, we usually set $t=1$.

By using this approach the length of the vectors $\vec{c}^{(n)}$ can be much smaller than the dimension of the Hilbert space 
while the expectation value of $\vec{c}^{(n)}$ still approaches the exact eigenvector of the ground state of $\mat{H}$. At the same time, the shift $S^{(n)}$ equilibrates to fluctuating around the ground state eigenvalue with a small stochastic bias \cite{Vigor2015,Brand2021}.
The spawning process described above differs from the original  one of Ref.~\cite{Booth2009} and is similar in spirit but more efficient than the modifications discussed in Refs.~\cite{Lim2015,Greene2019,Greene2020}. It will be described in greater detail elsewhere~\cite{Cufar2022ep}.

{While Eq.~\eqref{eq:2c-ham-mom} defines the Hamiltonian used in this study, the FCIQMC method is completely agnostic to the nature of the Hamiltonian, as long as it results in a sparse matrix where elements can be computed efficiently on the fly. Thus it is possible to study multi-dimensional models, long-range interactions, or even complex-valued problems \cite{Booth2013}.}

\subsection{Implementation Details}

We have implemented the FCIQMC algorithm in the high-level and high-performance programming language Julia \cite{Bezanson2017}. Both the FCIQMC algorithm and all analysis tools are implemented as a library. This way, calculations and all parameters can be defined in a concise script, written in the same language as the library, without the need for input files in a different format. The setup is  very flexible and makes it easy to experiment interactively with immediate visualization of data, e.g.~in a notebook interface, or deploy code to a high-performance computer. The library code  \texttt{Rimu.jl} used for all calculations in this work is available as an open-source software project  \cite{rimucode}.

While in practice the matrix $\mat{H}$ is extremely large, it is also extremely sparse and it is easy
to compute its matrix elements on the fly. To facilitate this, we index the matrix and vectors with the Fock states of Eq.~\eqref{eq:Fockstates} directly.
To encode the occupation number representation of a bosonic Fock
state, we use a bit string where a sequence of $n$ ones encodes $n$ particles in a mode (lattice site) and
zeros are used as separators between the modes. As an example, the state $| 0, 0, 3, 0, 1, 2
\rangle$ would be encoded as the bit string ``00111001011''.  Using this scheme, storing $N$
particles in $M$ modes requires a bit string of length $N + M - 1$. This representation is both extremely compact and allows for efficient on-the-fly calculations through bit manipulations. 

The \texttt{Rimu.jl} code makes extensive use of Julia's type system and code optimization capabilities through the multiple-dispatch paradigm and  just-in-time  compilation \cite{Bezanson2017}. E.g., the number of particles $N$ and modes $M$, and the length of a bitstring are all encoded in the type of a Fock-state address as type parameters. This allows us to easily write generic, well tested, and reusable library code for manipulating bit strings and matrix-element calculation for bit strings of arbitrary length and type.
As the type information is available at compile time, part of the computational workload related to specializing the code to a specific physical problem is off-loaded to the compiler.  
Julia's just-in-time compiler can thus produce optimized code for the particular parameters of the physics problem, which is easily defined in the script that is used to initiate the computation. As a consequence of this approach some lag from compilation is experienced in interactive use, but for the computationally intensive Monte Carlo calculations, the benefits from optimized code compilation are appreciable.



\subsection{Data structures and distributed Computation}

For representing the coefficient vector $\vec{c}$ it is important to access the data quickly based on the Fock space address. This is important as spawns hitting the same configuration must be allowed to annihilate~\cite{Booth2009}, but also to save memory by encoding all walkers on a single configuration in a single number. We thus use a dictionary data structure to store the non-zero elements of $\vec{c}$, which is realized as a hash table \cite{Booth2014} and thus provides access times that are nearly independent of the number on nonzero vector elements.


Another benefit of this approach is that it is relatively easy to distribute the data and computations to be processed in parallel.
Our approach to parallelization follows Ref.~\cite{Booth2014} and divides the vector $\vec{c}$ into
approximately equally-sized chunks, which are assigned  to different workers. The workers perform the
spawning step independently. After each step, but before the vector compression, a communication step
is performed, where the newly spawned entries are transferred to the correct workers.  In
our implementation, we use the Message Passing Interface (MPI)~\cite{mpi1993} through its Julia bindings \cite{Byrne2021} to handle the data distribution and communication between workers.

\subsection{The Initiator Approximation}

With some Hamiltonians, FCIQMC exhibits the sign problem. The problem manifests itself when
the number of walkers, which is equal to $\Vert\vec{c}\Vert_1$, is too small. In such a regime, the
energy estimates given by FCIQMC become completely unusable~\cite{Spencer2012}.

A well-known solution to the sign problem in FCIQMC is the initiator
approximation~\cite{Cleland2010}, which trades the sign problem for a small
bias. It works by suppressing spawns from configurations with low walker occupation. To be
precise, it divides the entries of $\vec{c}$ into two classes: initiators, and
non-initiators. For the initiators, the algorithm is unchanged, while the non-initiators are
only allowed to perform spawns to configurations that are themselves initiators. A
configuration is an initiator if its occupation number is strictly greater than a chosen
initiator threshold. In our computations, the initiator threshold was always set to 1.

\subsection{Simulation Details}\label{sec:simulation-details}

All simulations were performed with the  \texttt{Rimu.jl} \cite{rimucode} code (version \texttt{v0.6.0}) written by the authors.
Energy estimators are computed as averages from a time series collected from the simulation discarding data from an initial equilibration phase. The projected energy is used throughout this work as it has a much smaller fluctuation comparing to the shift estimator, provided sufficient number of walkers occupies the reference configurations.
Error bars were determined using the blocking analysis of Ref.~\cite{Flyvbjerg1989} supplemented by hypothesis testing of Ref.~\cite{Jonsson2018}.

When calculating an expectation value of an observable, such as the two-body correlation and the momentum of the impurity, the replica trick \cite{Overy2014} is used. It uses two independent FCIQMC wave functions to avoid a bias that would appear if correlated data was used.

For most of the calculations, one million floating point walkers are used. As mentioned previously, the initiator approach is applied to all calculations with a threshold value of $1$. This is necessary for controlling the sign problem in our simulations in the parameter regimes of larger values of $\gamma$ and $\eta$. We have performed extensive tests to control the biases introduced by population control and the initiator approximation and present some exemplary data from these efforts in App.~\ref{sec:Bias}.

For calculations with small $\eta$, the equilibration can take a very long time. To overcome this problem, we used equilibrated wave functions from a system with much larger $\eta$ as the starting vector, and re-equilibrated the wave function with the desired small $\eta$. This procedure speeds up the equilibration process significantly. 

%

\section{Results}\label{sec:results}

Yrast states are the lowest energy states at a given non-zero momentum. We denote the energy of the yrast state $|\Psi_P\rangle$ as $E_{N,N_\mathrm{imp}}(P)$ [or $E(P)$ for short], where $N$ is the particle number of the Bose gas, $N_\mathrm{imp}$ the number of impurities present, and $P$ the total (conserved) momentum. We also refer to the energy as a function of momentum as the ``dispersion''. In the thermodynamic limit where $N,L\to\infty$ while the density $n=N/L$ is finite, the momentum becomes a continuous variable.

\begin{figure}[h]
\includegraphics[width=10.5 cm]{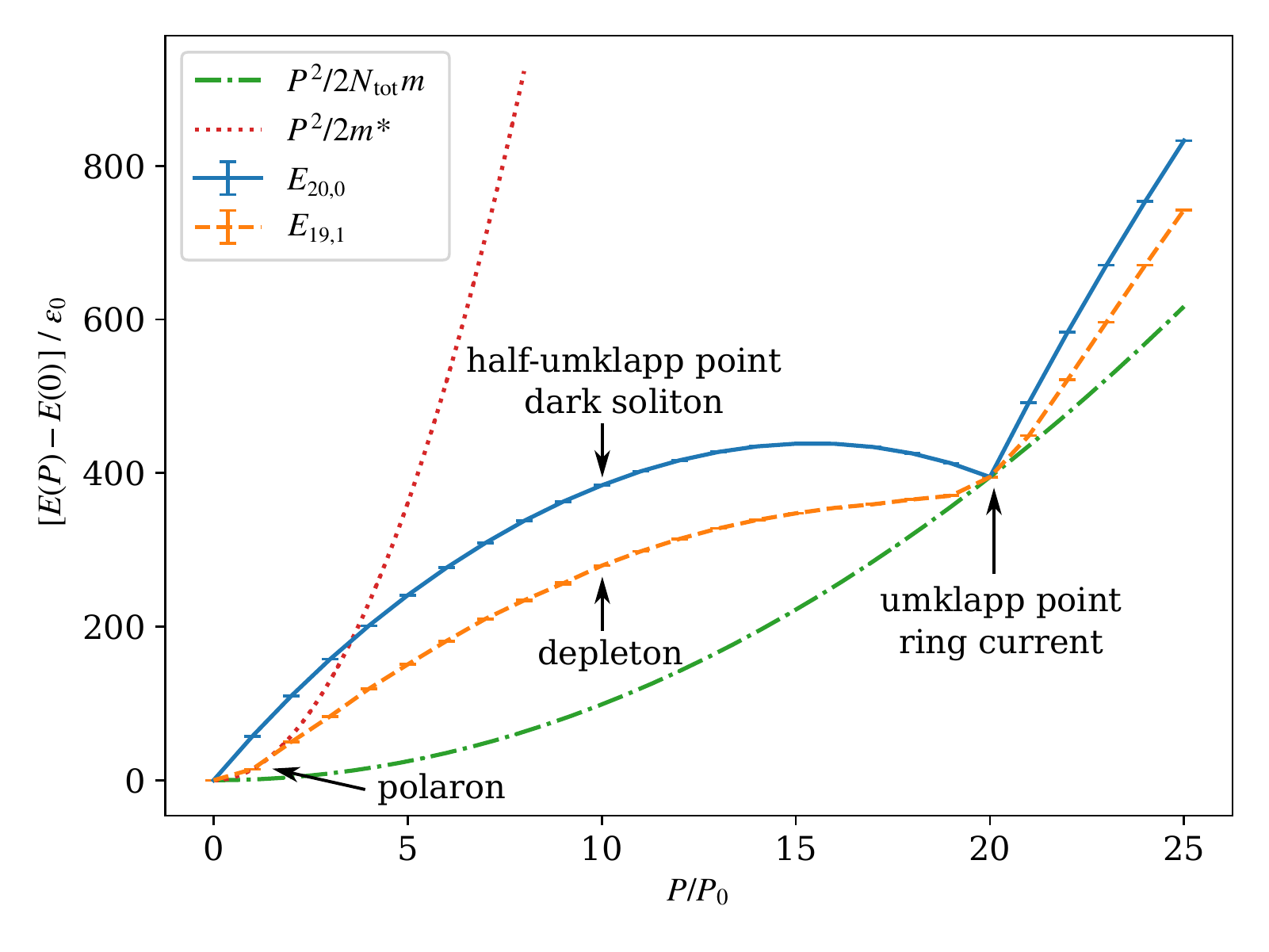}
\caption{\label{fig:yrast}
Yrast excitation energies of a finite Bose gas with $N_\mathrm{tot}=20$ particles with periodic boundary conditions.
The blue data (solid line as a guide to the eye) is for a weakly-interacting, pure Bose gas ($N=20$, $N_\mathrm{imp}=0$). A Bose gas containing a single spin impurity ($N=19$, $N_\mathrm{imp}=1$) with repulsive interactions ($\eta= 0.5$) is shown in  orange (dashed line as a guide to the eye). The interaction strength in the Bose gas is  $\gamma=0.2$.
All data points are obtained from FCIQMC calculations with a fixed systems size of $M=50$. 
The dash-dotted line shows the center-of-mass dispersion relation $E(P) - E(0) =  P^2/2N_\mathrm{tot}m$ for reference.
Finally, the dotted line depicts a quadratic polaron dispersion $E(P) - E(0) =  P^2/2m^*$, where $m^* = 1.3684(37) m$ is the fitted value of the polaron effective mass. The units of momentum and energy are $P_0 = 2\pi \hbar/L$ and $\varepsilon_0 = {\hbar^2}/{2mL^2}$.
}
\end{figure}

Yrast dispersions for a finite system with $N_\mathrm{tot} = N+N_\mathrm{imp} =20$ particles are shown in Fig.~\ref{fig:yrast}. Special points on the dispersion occur at integer multiples of the ``umklapp'' momentum $P = 2\pi \hbar N_\mathrm{tot}/L = N_\mathrm{tot} P_0$, where $P_0 = 2\pi \hbar/L$. At these umklapp points, the system's internal state is identical to the ground state with a Galilean boost applied, such that every particle gains a momentum of unit $P_0$. Thus, the umklapp points have the energy $E(P) = E(0) + P^2/(2N_\mathrm{tot}m)$, as indicated by the dash-dotted (green) parabola in Fig.~\ref{fig:yrast}.
  
For a pure one-dimensional Bose gas (where $N_\mathrm{imp}=0$), the yrast states and their energy can be generated exactly via the Bethe ansatz \cite{lieb63:2}\footnote{Yrast states in the one-dimensional Bose gas were denoted as type II excitations by Lieb in Ref.~\cite{lieb63:2}}. The yrast dispersion for $N=20$ bosons from FCIQMC is shown in Fig.~\ref{fig:yrast} by the blue data (solid line).
The umklapp points {at $P=N_\mathrm{tot}P_0$ (and integer multiples)} have the meaning of a superfluid ring current~\cite{Cherny2012}. 
The rest of the yrast dispersion are associated with dark and gray soliton phenomena  \cite{Tsuzuki1971} characterized by a localized dip in the density and step in the superfluid phase. While the momentum eigenstates are translationally invariant and can be thought of as a superposition of the (localized) solitons at various positions \cite{Fialko2012}, wave-packet-like superpositions of nearby momentum eigenstates reveal localized soliton solutions that move at the velocity of $v = dE(P)/dP$, given by the slope of the yrast dispersion \cite{Shamailov2019}. At the half umklapp momentum $P=N_\mathrm{tot} P_0/2$, a dark soliton forms with a $\pi$ phase step. It is associated with a negative effective mass $m^* = (d^2E/dP^2)^{-1}$, and as a consequence will oscillate around localized density maxima created by trapping potentials \cite{Konotop2004,Astrakharchik2012}. At small momentum (and next to any umklapp point), the Bose gas dispersion is linear and the slope becomes the Bogoliubov speed of sound in the thermodynamic limit~\cite{lieb63:2}. 

An yrast dispersion in the presence of a spin impurity ($N_\mathrm{imp}=1$) is shown with orange data (dashed line) in Fig.~\ref{fig:yrast}. The excitation energy of yrast states $E(P)-E(0)$ is generally smaller in the presence of a spin impurity compared to a Bose gas with the same number of particles $N_\mathrm{tot}$ apart from the umklapp points. This can be attributed to the fact that the spin impurity is not a part of the superfluid and thus does not fully contribute to the energy cost of forming a soliton by creating a twist in the phase -- a phenomenon that can be rationalized  with the phase rigidity of a superfluid \cite{Anderson1966}\footnote{Phenomena associated to phase rigidity occur in a one-dimensional Bose gas even though Bose-Einstein condensation is absent in the thermodynamic limit~\cite{Cherny2012,Shamailov2019}.}.
The yrast dispersion in the presence of the impurity is approximately quadratic at small momentum (and near the umklapp points), in contrast to the pure Bose gas, and thus can be assigned an effective mass. The fitted, idealized parabola is shown in Fig.~\ref{fig:yrast} as a dotted (red) line. The effective mass determined by the curvature may differ from the bare mass $m$ of the impurity due to interactions with the bosonic superfluid. We refer to this quadratic part of the dispersion near the umklapp points as the polaron. 

Near the half umklapp momentum at $P=N_\mathrm{tot} P_0/2$ we may expect the physics of a depleton, i.e.~a dark or gray soliton that is affected by the presence of the impurity \cite{Schecter2016}. 
{While Fig.~\ref{fig:yrast} depicts data for $N_\mathrm{tot}=20$ particles, the situation is generic (upon adjusting the scales) for finite particle number where the umklapp momentum is found at $P=N_\mathrm{tot} P_0$.}

\subsection{Yrast dispersion with weak and strong boson-impurity coupling strength} \label{sec:yrast_disp}

Figure~\ref{fig:yrast} displays strong finite size effects in terms of the center-of-mass dispersion ${P^2}/{2N_\mathrm{tot}m}$, the classical kinetic energy associated with the translation of the whole system, which provides a lower limit for the yrast excitation energies (shown as a dash-dotted line). In the thermodynamic limit this energy contribution vanishes due to the total system mass appearing in the denominator. The detailed relation between the yrast dispersion of a finite system and its thermodynamic limit has been worked out for the pure Bose gas in  Ref.~\cite{Shamailov2019} in terms of quantities like the phase step, the associated superfluid backflow current, and the depleted particle number for a dark/gray soliton. Here we 
correct for the dominant finite size effect in a simple way
by subtracting the center-of-mass kinetic energy from the yrast excitation energy. 
We thus define the finite-size corrected yrast dispersion, $\Omega(P)$, as
\begin{equation}\label{eq:Omega}
    \Omega(P) = E(P) - E(0) - \frac{P^2}{2N_\mathrm{tot}m},
\end{equation}
where $E(P)$ is the lowest energy at fixed momentum $P$. 
The finite-size energy correction is equivalent to a Galilean boost into a reference frame that moves with the velocity $P/{N_\mathrm{tot}m}$ that a classical particle of mass ${N_\mathrm{tot}m}$ would have at momentum $P$.
By removing the center of mass kinetic energy from the total energy, the yrast dispersion becomes periodic in $P$ with the umklapp momentum $2\pi \hbar N_\mathrm{tot}/L = N_\mathrm{tot} P_0$ as the period, as in the thermodynamic limit. Additionally,  the finite size corrected yrast dispersion has reflection symmetry across the half-umklapp point $N_\mathrm{tot} P_0/2$. 

\end{paracol}
\begin{figure}[htb]
    \widefigure
    \begin{tabular}{cc}
        (a) & (b) \\
        \includegraphics[height=6.5 cm]{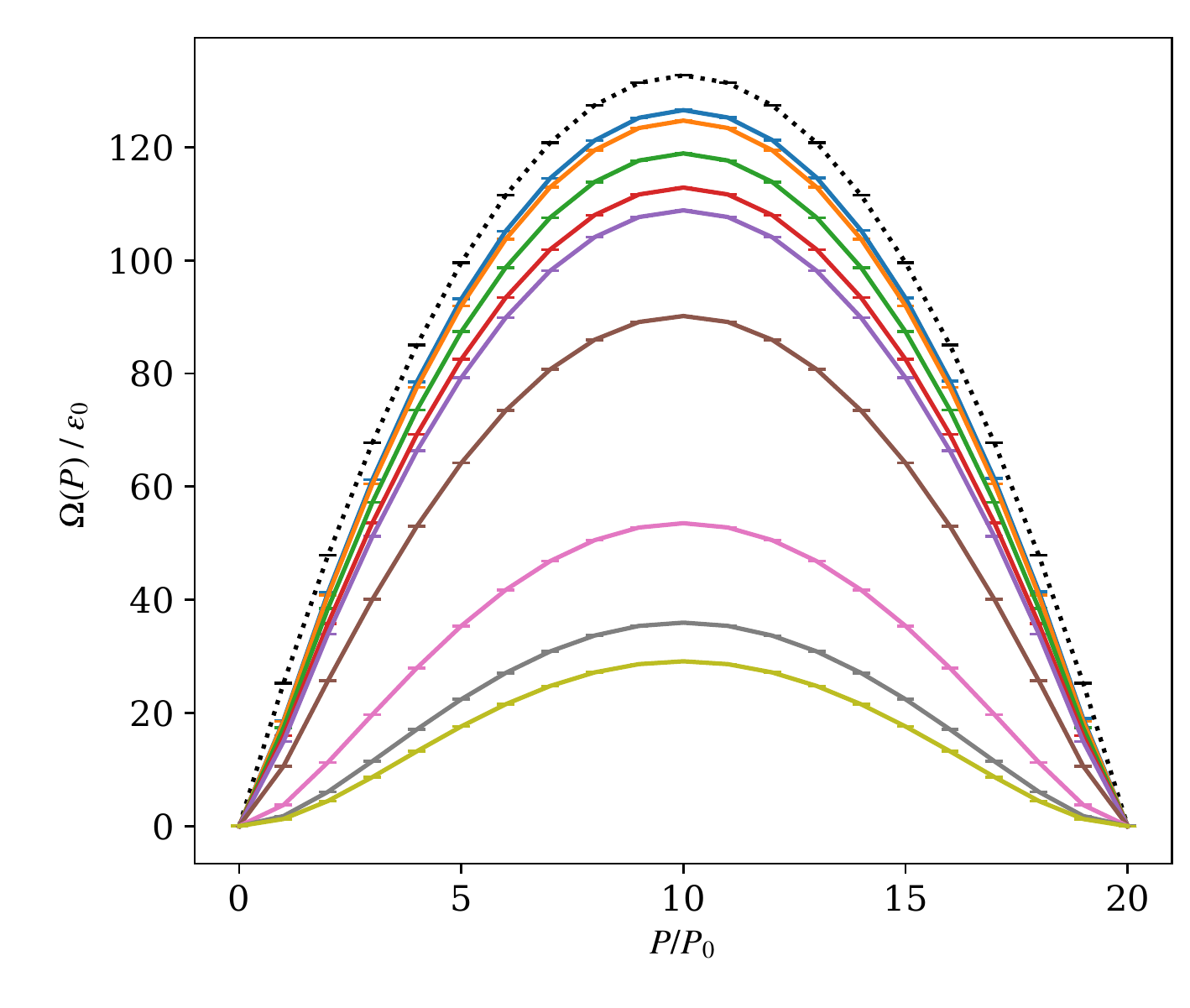} & \includegraphics[height=6.5 cm]{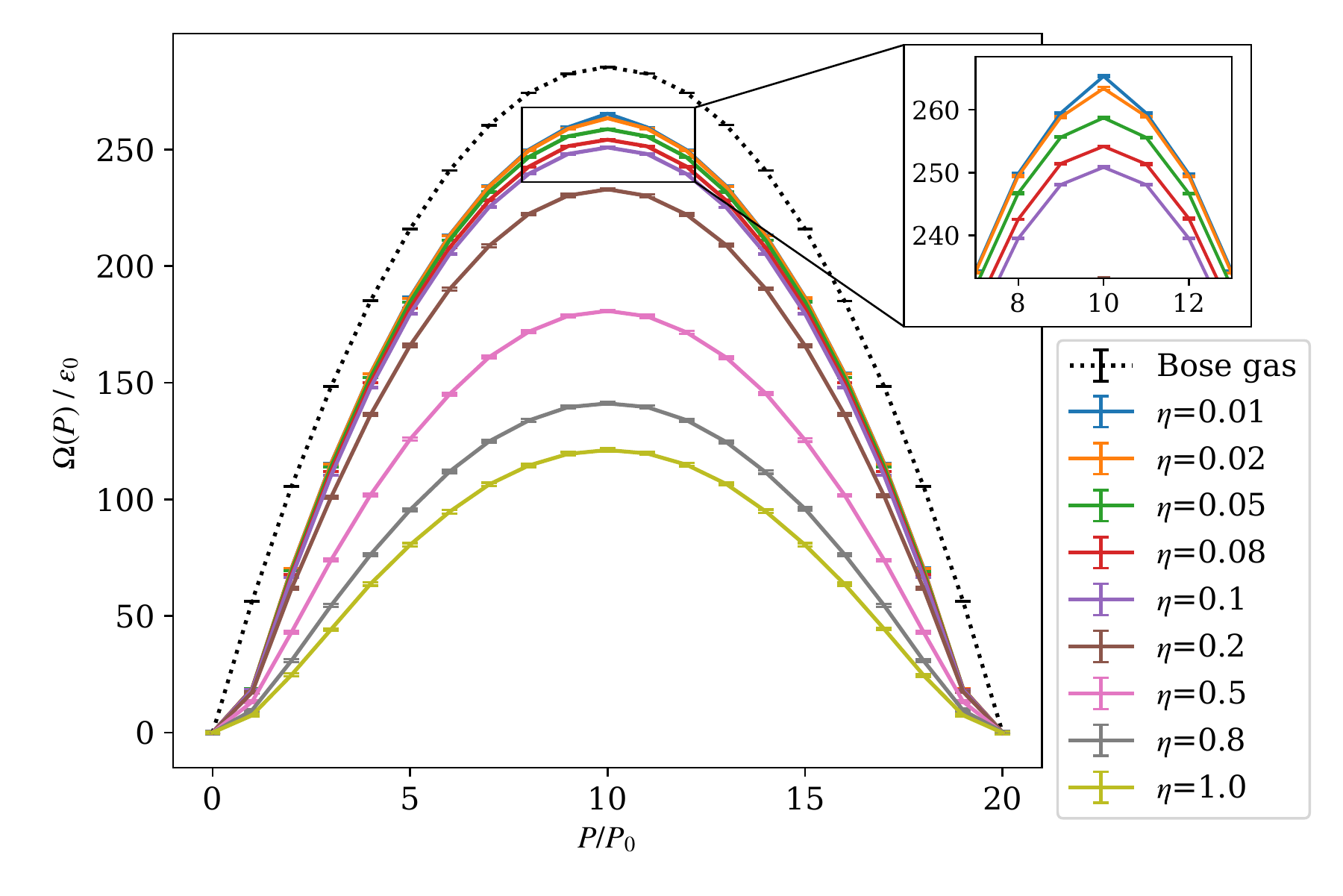}
    \end{tabular}
    \caption{\label{fig:yrast-gamma}Finite-size-corrected yrast dispersion of a pure Bose gas ($N=20$, $N_\mathrm{int}=0$, black markers with dotted line) and Bose gas with spin impurity   ($N=19$, $N_\mathrm{int}=1$, colored markers) according to Eq.~\eqref{eq:Omega}. The boson-boson coupling strength is (a) $\gamma=0.02$ and (b) $\gamma=0.2$. FCIQMC results are shown with error bars for fixed systems size $M=50$. The boson-impurity coupling strength is varied between $\eta = 0.01$ and $\eta = 1.0$ as per legend. The pure Bose gas excitation energies are higher than any of the impurity dispersion data.
Note that a slight cusp develops in the impurity dispersion at the half-umklapp point, ${P}=10{P_0}$ in the regime where $\eta \ll \gamma$ in panel~(b) {as can be seen from the inset, which magnifies the data in the region of the cusp}. }
\end{figure}
\begin{paracol}{2}
\switchcolumn

In Fig.~\ref{fig:yrast-gamma}, we present two sets of finite-size corrected yrast dispersions with boson-boson coupling strengths of $\gamma=0.02$ and $\gamma=0.2$, which are both considered to be weak interactions. The boson-impurity coupling is chosen in the range from $\eta = 0.01$ to 1.0, which covers both $\eta > \gamma$ and $\eta < \gamma$ scenarios.
Our results show that the yrast excitation energy is consistently lower in the presence of the spin impurity compared to the pure Bose gas at any value of $\eta$, as previously predicted~\cite{Lamacraft2009,Schecter2016}.
The quadratic polaron part of the dispersion near $P=0$ and the umklapp points reduces its curvature with increasing $\eta$, which is consistent with an increase of the polaron effective mass. Quantitative results for the polaron effective mass were previously reported from diffusion Monte Carlo calculations  \cite{Parisi2017} and mean-field theory \cite{Panochko2019}. 

%

As shown in Fig.~\ref{fig:yrast-gamma}, the shape of the yrast dispersion is smooth in general. However, although very subtle, a cusp at the half-umklapp point can be seen in panel~(b) when $\eta\ll\gamma$. 
A cusp for weak coupling in an infinite system was predicted in Ref.~\cite{Lamacraft2009} based on Luttinger liquid theory. It was pointed out that in a Luttinger liquid the cusp is expected to vanish discontinuously at some critical value of the coupling strength between the impurity and the quantum liquid, but the exact transition point is difficult to determine. 
While we have only finite system data available from our calculations, we examine this transition in the remainder of this work and present further data that provides insights into the physics at play. 
%
%
%

\begin{figure}[htb]
    \centering
    \includegraphics[height=6.5 cm]{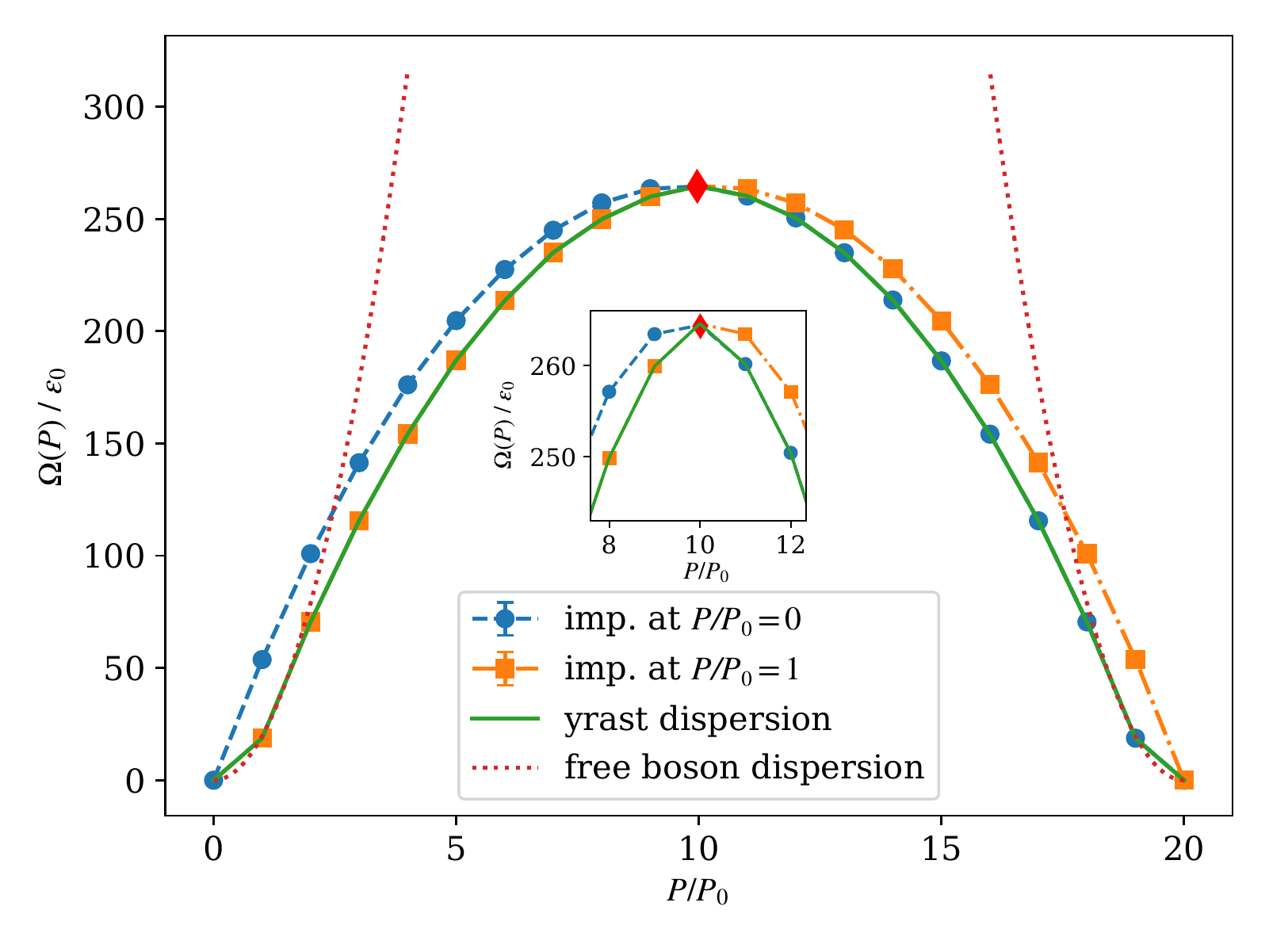}
    \caption{\label{fig:yrast-NI} Dispersion of a Bose gas with boson-boson coupling strength $\gamma=0.2$ and a single impurity with no coupling ($\eta=0$). The total number of particles is $N_\mathrm{tot}=20$. The impurity is given zero ($\bullet$) or one ($\blacksquare$ ) unit of momentum. The solid line shows the actual yrast dispersion with lowest energy states exhibiting a cusp at the half-umklapp point ($P=10P_o$) marked by the diamond. The dotted lines show the free particle dispersion $P^2/2m_\mathrm{free}$. The free particle mass differs slightly from the bare boson mass $m$ due to the finite-size correction in Eq.~\eqref{eq:Omega}. {The inset shows the detail of the cusp near the half-umklapp point ($P=10P_o$).}
    }
\end{figure}

It is easiest to understand the origin of the cusp from the situation where the impurity does not interact with the Bose gas at all.
We thus show data for 
a non-interacting impurity ($\eta=0$) immersed in a weakly interacting Bose gas at $\gamma=0.2$ in Fig.~\ref{fig:yrast-NI} , which demonstrates two interesting transition points in the yrast dispersion.
The transition points originate from a trade-off between the energy cost of depositing momentum into either the impurity or the Bose gas.
While for small momentum it is favorable to deposit momentum into the impurity (dotted red line), at $P>P_0$ it becomes favorable to deposit additional momentum into the Bose gas instead (orange squares show data where $P_\mathrm{imp}=P_0$). This is the first transition point. The second transition happens at the half umklapp point (red diamond), where the yrast state (indicated with a green line) switches again to one with zero impurity momentum $P_\mathrm{imp}=0$. 
This switch generates a cusp in the yrast dispersion, connecting segments with a gray soliton moving to the right (at $P<10P_0$) and a gray soliton moving to the left ($P>10P_0$).

A symmetrical scenario to the first transition happens near the umklapp point. 
Both transitions become sharp quantum phase transitions (cusp with discontinuous derivative) in the thermodynamic limit. 
{These phase transitions are first order (level-crossing type) transitions without diverging quantum fluctuations.} 

\subsection{Impurity momentum} \label{sec:Pimp}

In order to understand the physical nature of an yrast state   it is of great interest to understand how the momentum is distributed between the impurity and the Bose gas. In the case of an interacting impurity, it's momentum is no longer a good quantum number. Thus, we calculate the expectation value of the impurity momentum with respect to the yrast state $|\Psi_P\rangle$ 
\begin{equation}
    \langle \hat P_\mathrm{imp} \rangle_P =  \sum_{k=k_1}^{k_M} k P_0 \langle\Psi_P| \hat b^\dag_k \hat b_{k}|\Psi_P\rangle .
\end{equation}
Unbiased estimators for such a symmetric expectation value can be obtained from FCIQMC using the replica trick: two propagating independent stochastic representations of the quantum state are used for the bra and the ket state respectively \cite{Overy2014}. 

Figure~\ref{fig:mom-imp} shows the expectation value of the impurity momentum $\langle \hat P_\mathrm{imp} \rangle_P$ as a function of the total momentum of the yrast state for different values of the impurity coupling strength $\eta$ as blue dots. Because the total momentum is fixed to the value $P$ for each state, the expectation value of momentum in the Bose gas is given by the difference $\langle \hat P_\mathrm{Bg} \rangle_P = P - \langle \hat P_\mathrm{imp} \rangle_P$, where $\hat P_\mathrm{Bg} =  \sum_{k=k_1}^{k_M} k P_0 \hat a^\dag_k \hat a_{k}$ is the operator for the momentum of the Bose gas alone.

\end{paracol}
\begin{figure}[hp]
    \widefigure
    \begin{tabular}{cc}
        (a) $\eta=0.01$ & (b) $\eta=0.05$ \\
        \includegraphics[height=6.5 cm]{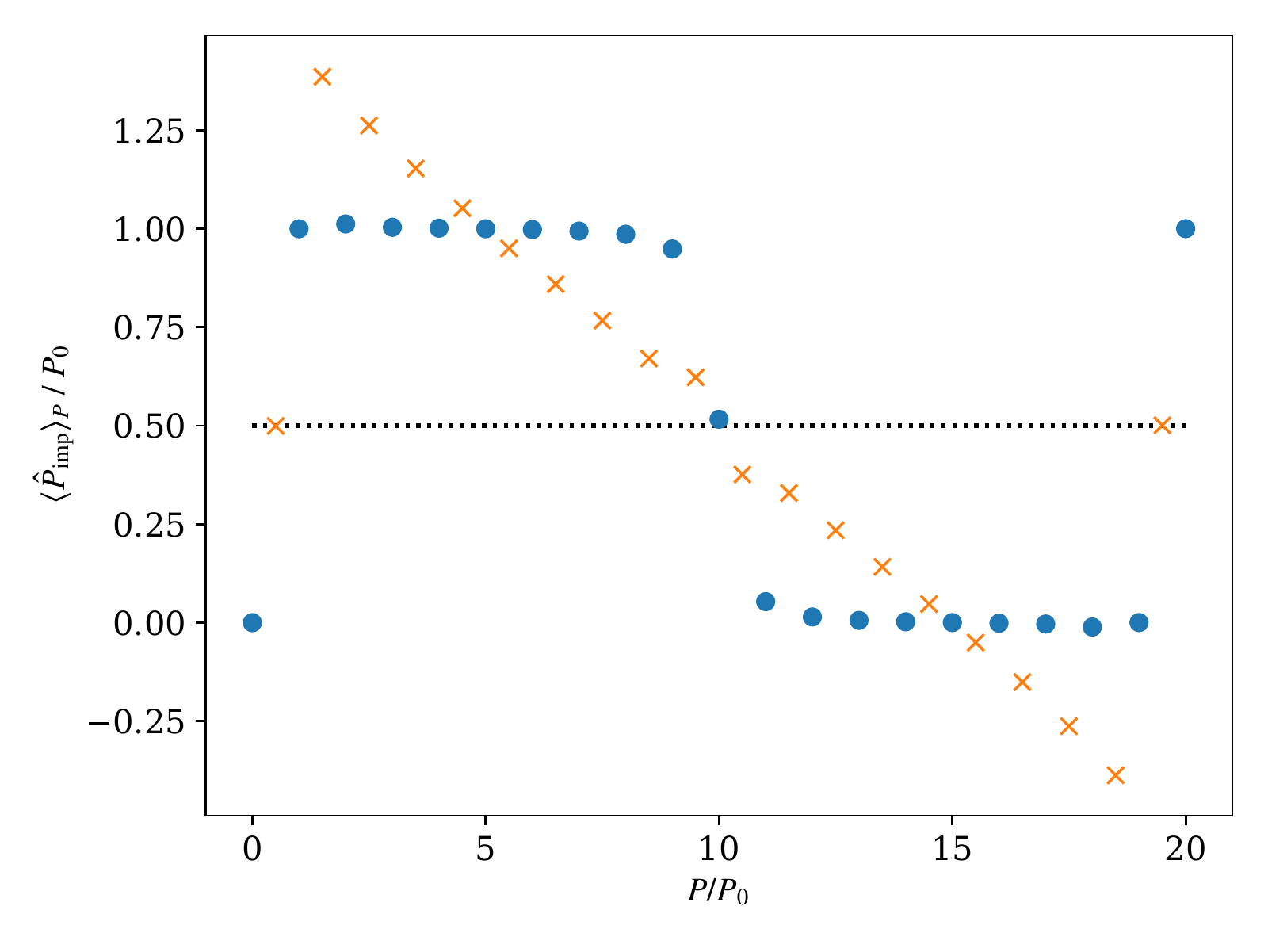} & \includegraphics[height=6.5 cm]{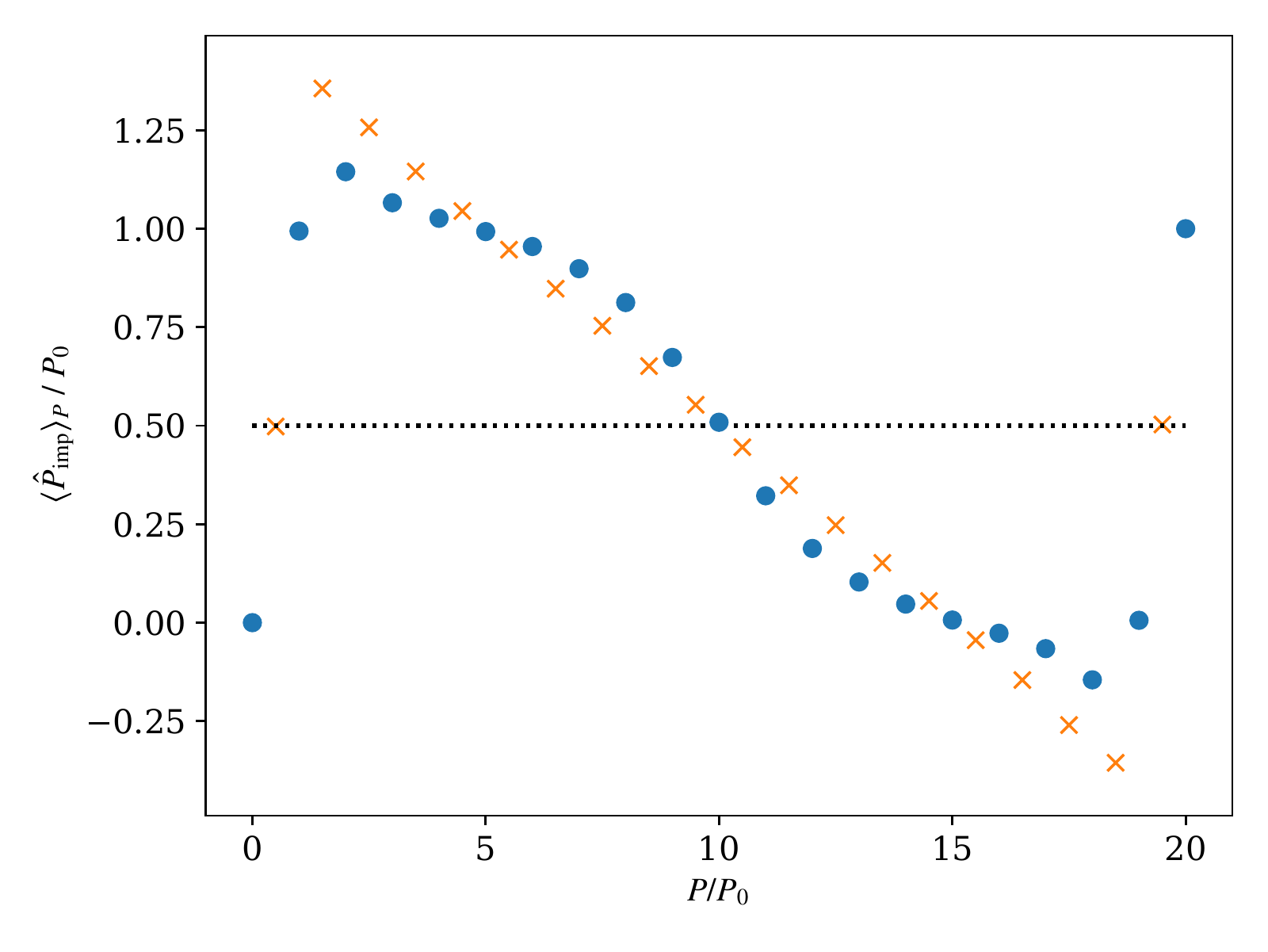}\\
        (c) $\eta=0.2$ & (d) $\eta=1$ \\
        \includegraphics[height=6.5 cm]{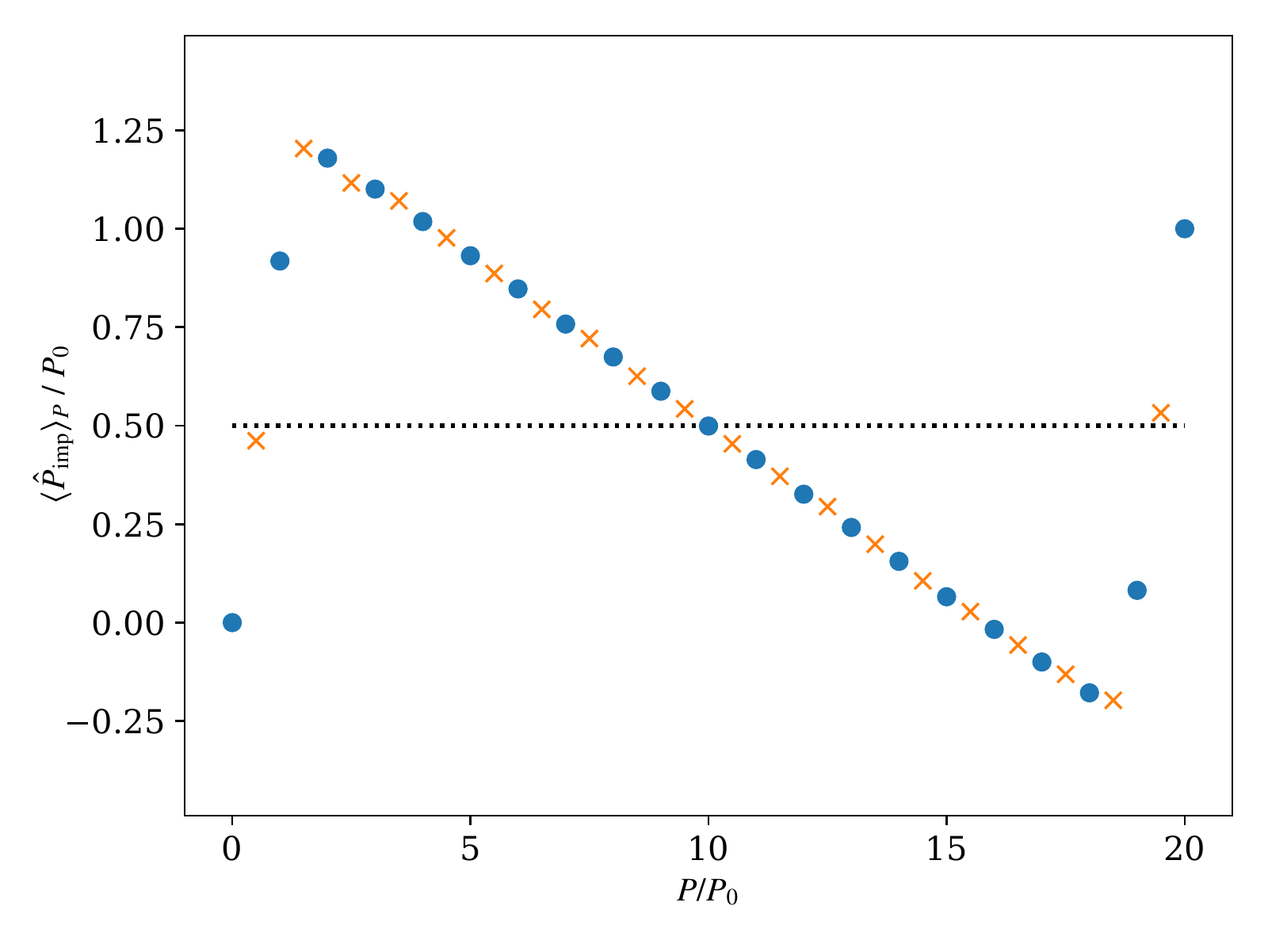} & \includegraphics[height=6.5 cm]{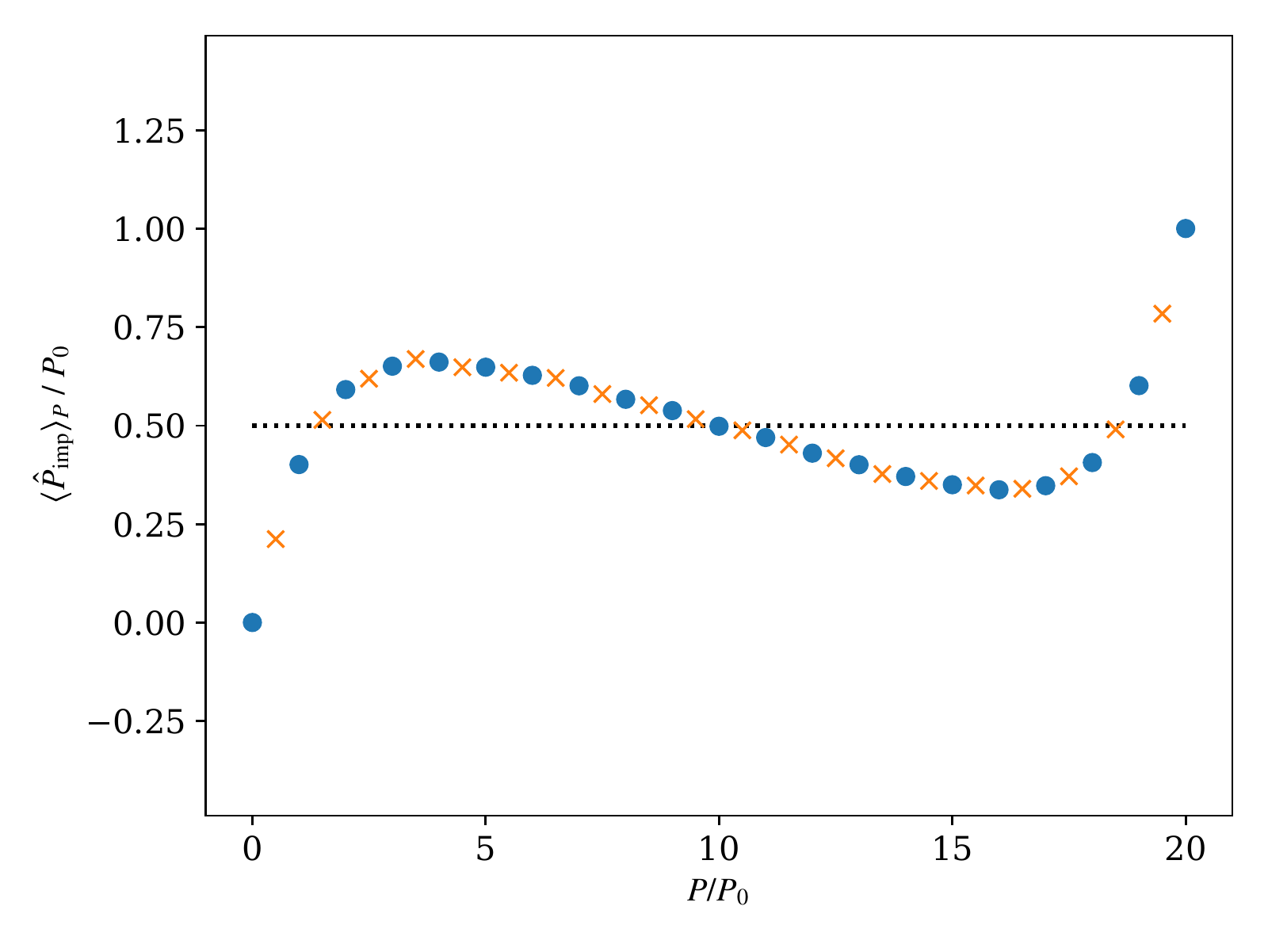}
    \end{tabular}
    \caption{\label{fig:mom-imp} The expectation value of the impurity momentum  $\langle \hat P_\mathrm{imp} \rangle_P$ against the total momentum $P$ in the system. The dots ($\bullet$) are directly calculated data with FCIQMC. The crosses ($\times$) show the $m\frac{dE}{dP}$ computed numerically from the yrast spectum in Fig.~\ref{fig:yrast-gamma}. 
    The dotted line shows the  value $\langle \hat P_\mathrm{imp} \rangle_P = 0.5 P_0$ as a guide to the eye.  The boson-boson coupling is $\gamma=0.2$ for all cases, and $N=19$ and  $N_\mathrm{imp}=1$, which means that $P = 10P_0$ corresponds to half umklapp and   $P = 20P_0$ is the full umklapp point.}
\end{figure}
\begin{paracol}{2}
\switchcolumn



In the case of weak impurity coupling ($\eta=0.01$) shown in Fig.~\ref{fig:mom-imp}(a) the situation is very close to the non-interacting limit of Fig.~\ref{fig:yrast-NI} discussed in the previous section: For small total momentum $P=P_0$ the impurity carries (almost) the full momentum of the system as this is energetically favourable. At larger values of $P$, additional momentum is taken up by the Bose gas while the impurity momentum stays at about $P_0$, before switching abruptly to approximately zero at the half umklapp point. At the full umklapp point $P=20 P_0$ the impurity momentum jumps back to $P_0$ consistent with the expectation that every particle including the impurity carries a single unit of quantised momentum at the umklapp point. The abrupt change near the half-umklapp point $P=10P_0$ is consistent with the cusp observed in the yrast dispersion in Fig.~\ref{fig:yrast-gamma}(b). An interesting situation occurs directly at the half umklapp point where $\langle \hat P_\mathrm{imp} \rangle\approx 0.5 P_0$, which indicates that this state is an entangled superposition of a state where the impurity has momentum $P_0$ and the Bose gas $9P_0$,  and a state with  $\langle \hat P_\mathrm{imp} \rangle\approx 0$ where the Bose gas carries the full (half-umklapp) momentum $10P_0$. 

At larger interaction strengths $\eta$ shown in panels (b) to (d), the curves keep the inversion symmetry around the half-umklapp point. At this point we find the entangled superposition state as described with $\langle \hat P_\mathrm{imp} \rangle\approx 0.5 P_0$. Strong changes are found in the polaron regions. At intermediate interactions additional momentum is deposited in the impurity with a maximum of $\langle \hat P_\mathrm{imp} \rangle\approx 1.25$ at $\eta=\gamma=0.2$. Further increase of the  impurity coupling then leads to a reduced expectation value for the impurity momentum going against $\langle \hat P_\mathrm{imp} \rangle\approx 0.5 P_0$ over the whole $P$ range.

The panels of Fig.~\ref{fig:mom-imp} also show  $m\frac{dE}{dP}\equiv mv$ where $v$ is the group velocity of the system with orange crosses. This data indicates that the impurity moves with the group velocity in the polaron part of the dispersion relation (close to $P=0$ or umklapp points) for a weakly-interacting impurity, and over the whole dispersion relation when $\eta\gtrapprox\gamma$. When $\eta\ll\gamma$ and outside of the polaron section, the impurity is rather transparent to the Bose gas and does not follow the group velocity. 
This indicates that the depleton picture where the impurity hybridizes with a dark or gray soliton is only valid when $\eta\gtrapprox\gamma$.

%
%


\subsection{Two-body correlation function} \label{sec:g2}

\end{paracol}
\begin{figure}[hp]
    \widefigure
    \begin{tabular}{cc}
        (a) $\eta=0.01$ & (b) $\eta=0.05$ \\
        \includegraphics[height=6.5 cm]{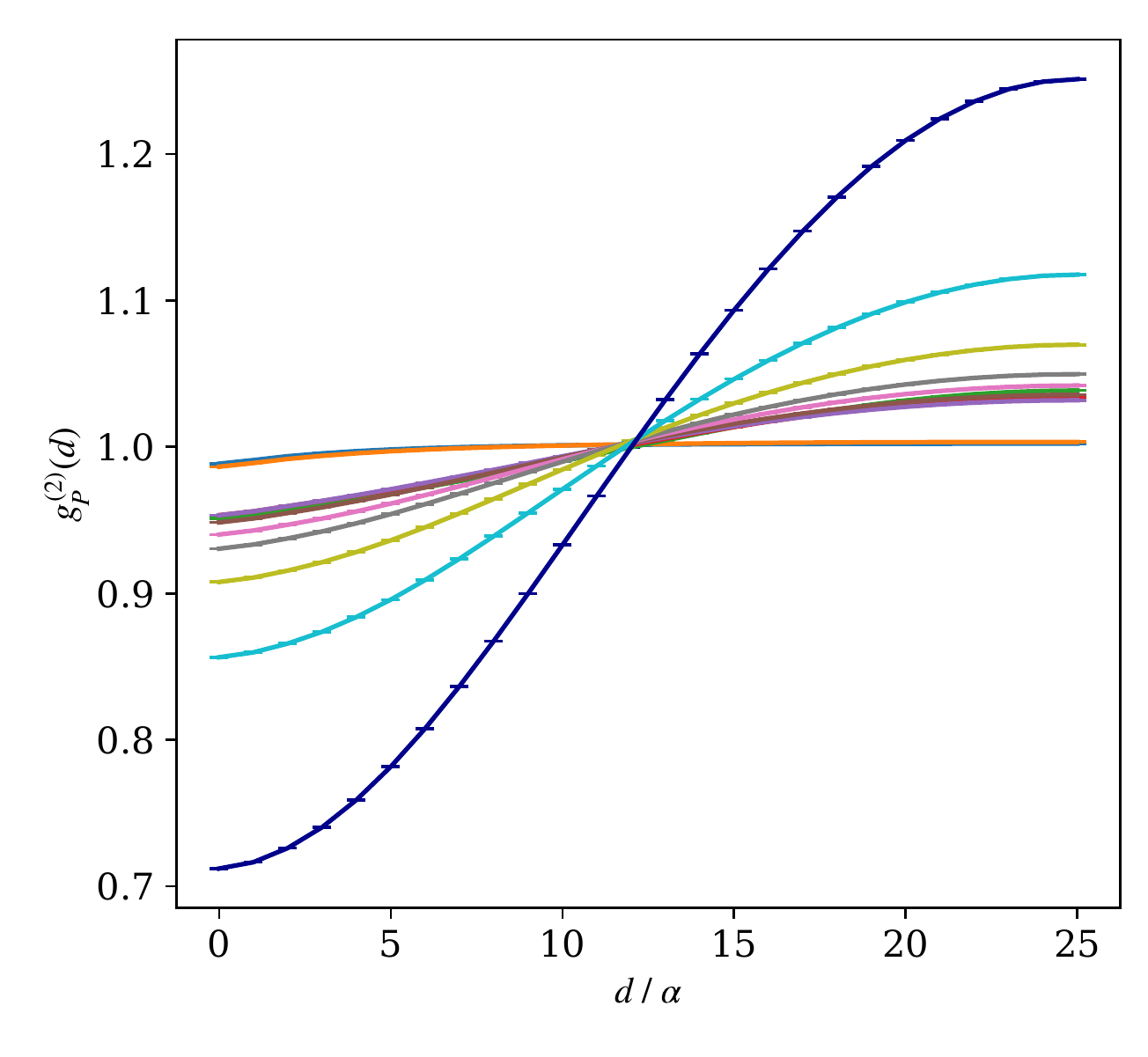} & \includegraphics[height=6.5 cm]{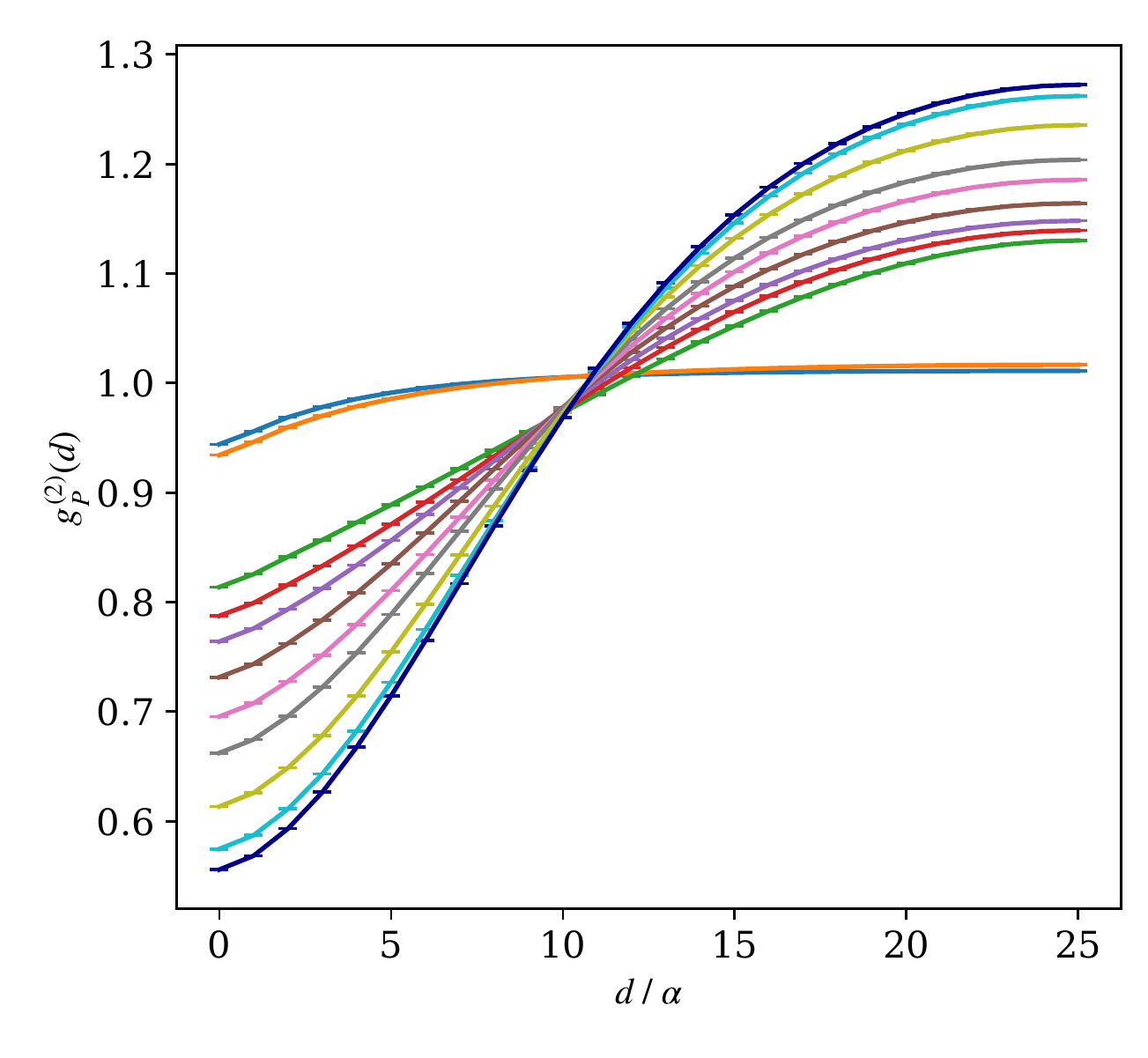}\\
        (c) $\eta=0.2$ & (d) $\eta=1$ \\
        \includegraphics[height=6.5 cm]{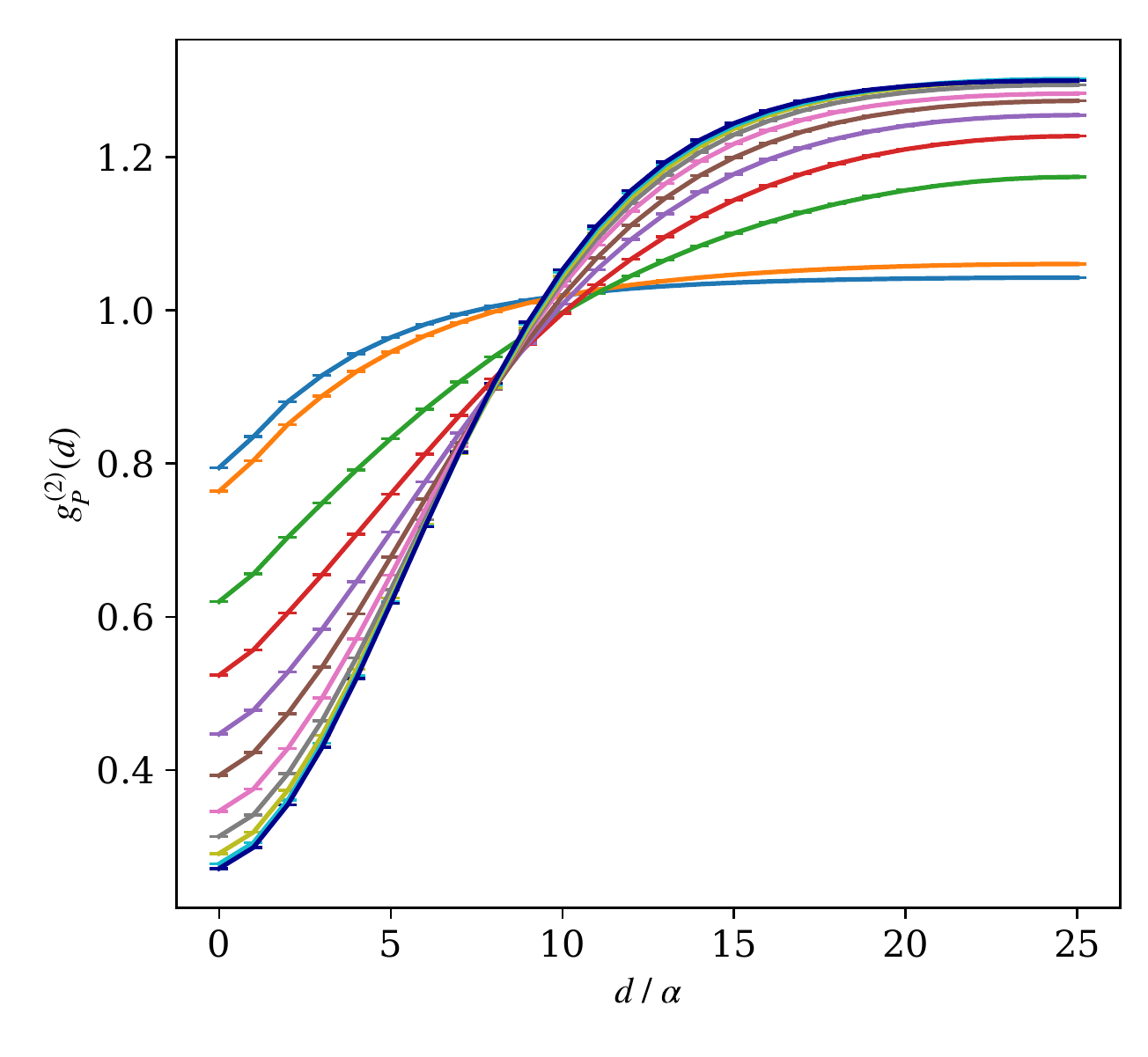} & \includegraphics[height=6.5 cm]{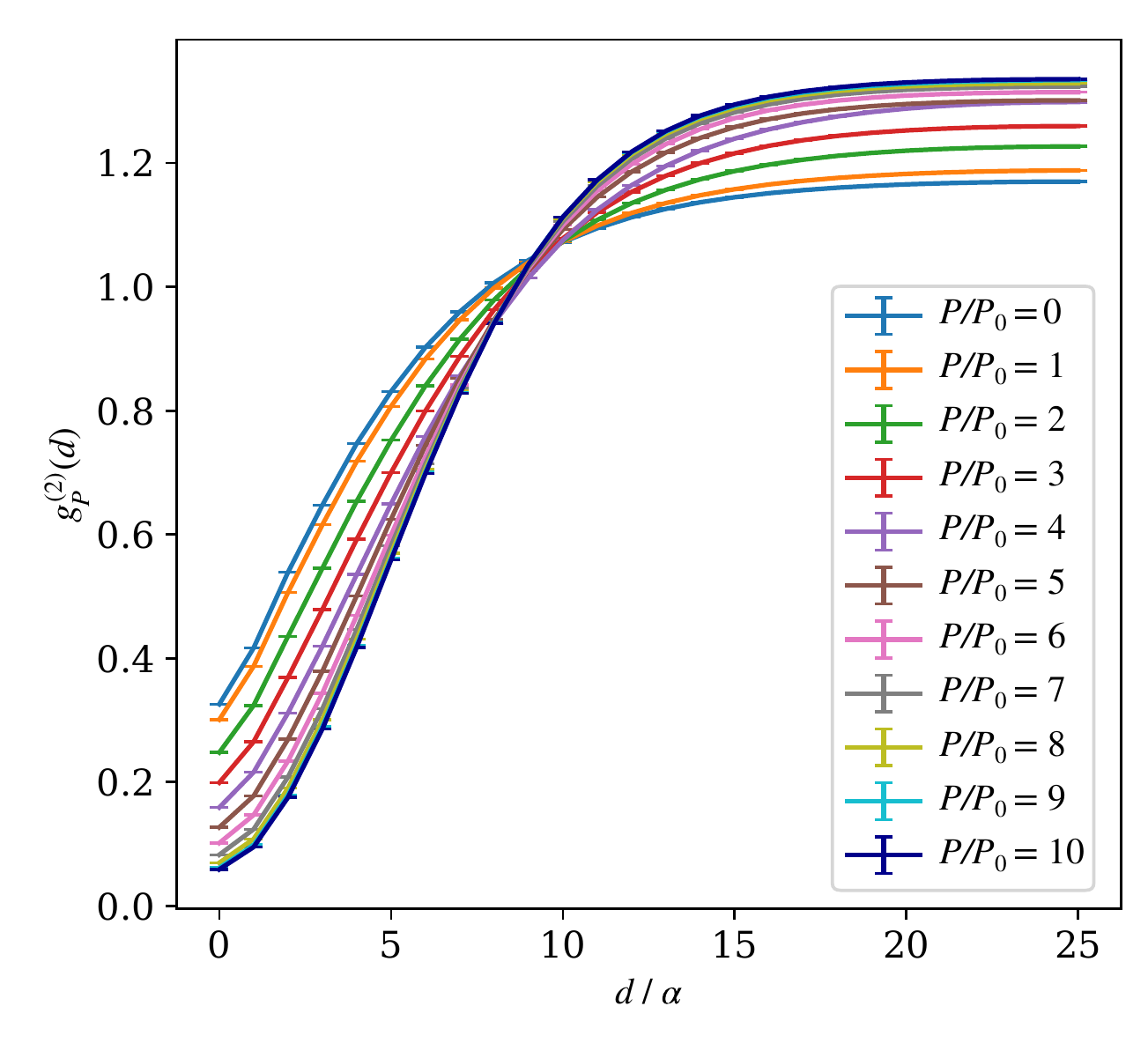}
    \end{tabular}
    \caption{\label{fig:g2-gamma02}The impurity-boson correlation function $g^{(2)}_P(d)$ for yrast states with total momentum $P$ as indicated in the legend versus the real space distance $d$. Different values of the impurity-boson coupling are shown in panel (a) $\eta=0.01$,  (b) $\eta=0.05$, (c) $\eta=0.2$ and (d) $\eta=1$. The boson-boson coupling is $\gamma=0.2$ for all cases, and $N=19$ and  $N_\mathrm{imp}=1$, which means that $P = 10P_0$ corresponds to half umklapp. 
    }
\end{figure}
\begin{paracol}{2}
\switchcolumn

The two-body correlation function contains important information about how particles interact with one another. In particular, the impurity-boson correlation provides direct evidence for the transition from a polaron to a depleton. Here, we define the dimensionless impurity-boson correlation function $g^{(2)}_P(d)$ for the yrast state $\ket{\Psi_P}$ in real space as
\begin{equation}\label{eq:g2-real}
g^{(2)}_P(d) =\frac{L}{N}\int_0^L \bra{\Psi_P} \hat\psi^\dag(x+d) \hat\psi_\mathrm{imp}^\dag(x)\hat\psi_\mathrm{imp}(x)\hat\psi(x+d)\ket{\Psi_P}\,\mathrm{d}x ,
\end{equation}
where $d$ is the distance between the impurity and a boson.  In order to evaluate this correlation function in the lattice discretized model we transform into momentum space using $\hat a_k^\dag = \int e^{ikx/\alpha}\hat\psi^\dag(x)\, \mathrm{d}x$ and $\hat b_k^\dag = \int e^{ikx/\alpha}\hat\psi_\mathrm{imp}^\dag(x)\, \mathrm{d}x$, to obtain the equivalent representation
\begin{align}
g^{(2)}_P(d) =  \frac{1}{M}  \sum_{s,p,q,r=1}^M  \exp({-id(p-q)\frac{2\pi}{L}}) \bra{\Psi_P} \hat{a}_s^\dag \hat{b}_p^\dag \hat{b}_q \hat{a}_r \ket{\Psi_P} \delta_{s+p,q+r}.
\end{align}
The chosen normalization ensures  $g^{(2)}_P(d)=1$ in a non-interacting system for any yrast state.
In an interacting system $g^{(2)}_P$ still obeys a reflection symmetry $g^{(2)}_P(d)$ = $g^{(2)}_P(-d)$ and is a periodic function with period $L$. Furthermore, as a function of the yrast momentum $P$, the correlation function $g^{(2)}_P$ of a finite system is periodic in $P$ with reflection symmetry around $P=0$ and around the half-umklapp point, as does the yrast dispersion relation in the thermodynamic limit. Due to these symmetries we show  the correlation functions only in the nontrivial intervals $0\le d\le L/2$ and $0\le P \le N_\mathrm{tot} P_0/2$.
%
%

Figure~\ref{fig:g2-gamma02} shows the correlation functions $g^{(2)}_P(d)$ for yrast states with different momentum  over a range of impurity-boson coupling strengths for  $\gamma=0.2$. 
Significant changes in the correlation functions with respect to different momentum values are clearly visible.

For the smallest interaction strength $\eta=0.01$ in Fig.~\ref{fig:g2-gamma02}(a) we can identify clear evidence of the two transitions discussed in Sec.~\ref{sec:yrast_disp}: For $P \le P_0$ the very small deviations of $g^{(2)}_P(d)$ from the background value of 1 indicate very weak correlations consistent with the polaron regime. There is evidence for a weak correlation hole, and the shape of $g^{(2)}_P(d)$ (negative curvature) is consistent with an otherwise homogeneous Bose gas. In the intermediate momentum range $P_0 < P < 10P_0$ the correlations are stronger and the shape of $g^{(2)}_P(d)$ changes with displaying positive curvature at small $P$ to negative at larger $P$. This is consistent with the impurity weakly correlating with a gray soliton forming in the Bose gas -- explaining the shape of the correlation function. A much stronger correlation is observed at the half umklapp point $P=10P_0$ consistent with the superposition state expected at the cusp of the dispersion as discussed in Sec.~\ref{sec:Pimp}.

Increasing the impurity-boson coupling strength $\eta$ in  panels~\ref{fig:g2-gamma02}(b)--(d) the changes in the correlation function for different $P$ values become smaller. For $\eta=0.05$ in panel (b) the transition from the half-umklapp momentum $P=10P_0$ to smaller momentum values is less dramatic and smoother, which indicates that the depleton picture of the impurity being localized inside a (modified) gray soliton is becoming adequate. However, comparing with Fig.~\ref{fig:mom-imp}(b) we see that this is not yet completely the case and still requires larger $\eta$ to become fully accurate.

The physics of the polaron regime is more resilient and survives to larger values of $\eta$ up to $\eta\approx 0.2$ as seen in panels~\ref{fig:g2-gamma02}(b) and (c). We note that the impurity carries almost the full momentum of the system at $P=P_0$ for $\eta\le 0.2$ as seen in Fig.~\ref{fig:mom-imp}, which is consistent with the polaron picture. Seeing the (anti-)correlation with the Bose gas strengthened with increasing $\eta$ in panels~\ref{fig:g2-gamma02}(b) and (c) is consistent with the decrease in the curvature of the dispersion observed in Fig.~\ref{fig:yrast-gamma} and associated increase in polaron mass.

At the largest value of $\eta=1$ shown in Fig.~\ref{fig:g2-gamma02}(d) the correlation function drops close to zero at zero distance $d=0$ for any value of $P$ consistent with the picture that the impurity now acts as a weak link in the Bose gas, almost severing the superfluid \cite{Schecter2016}. At the half umklapp momentum $P=10P_0$ the shape of the correlation function now closely traces the shape of a dark soliton density $\sim \tanh(d/l_\mathrm{h})^2$, consistent with a healing length $l_\mathrm{h} \equiv L/\sqrt{2\gamma}N \approx 4.2 \alpha$.

%

\subsection{Effective mass at half umklapp $P=N_\mathrm{tot} P_0/2$.} \label{sec:mass}

At the half umklapp point $P=N_\mathrm{tot} P_0/2$ we may expect the physics of the yrast states to be dominated by a dark soliton in the interacting Bose gas, and by a depleton if an interacting spin impurity is present. The effective mass $m^* = (d^2E/dP^2)^{-1}$ is negative due to the concave shape of the dispersion relation. We extract the effective mass by fitting a parabola to three points of the finite-size corrected dispersion relation $\Omega(P)$ of Eq.~\eqref{eq:Omega} near the half umklapp point.

%

\end{paracol}
\begin{figure}[htb]
    \widefigure
    \begin{tabular}{cc}
        (a) & (b) \\
\includegraphics[height=6.5 cm]{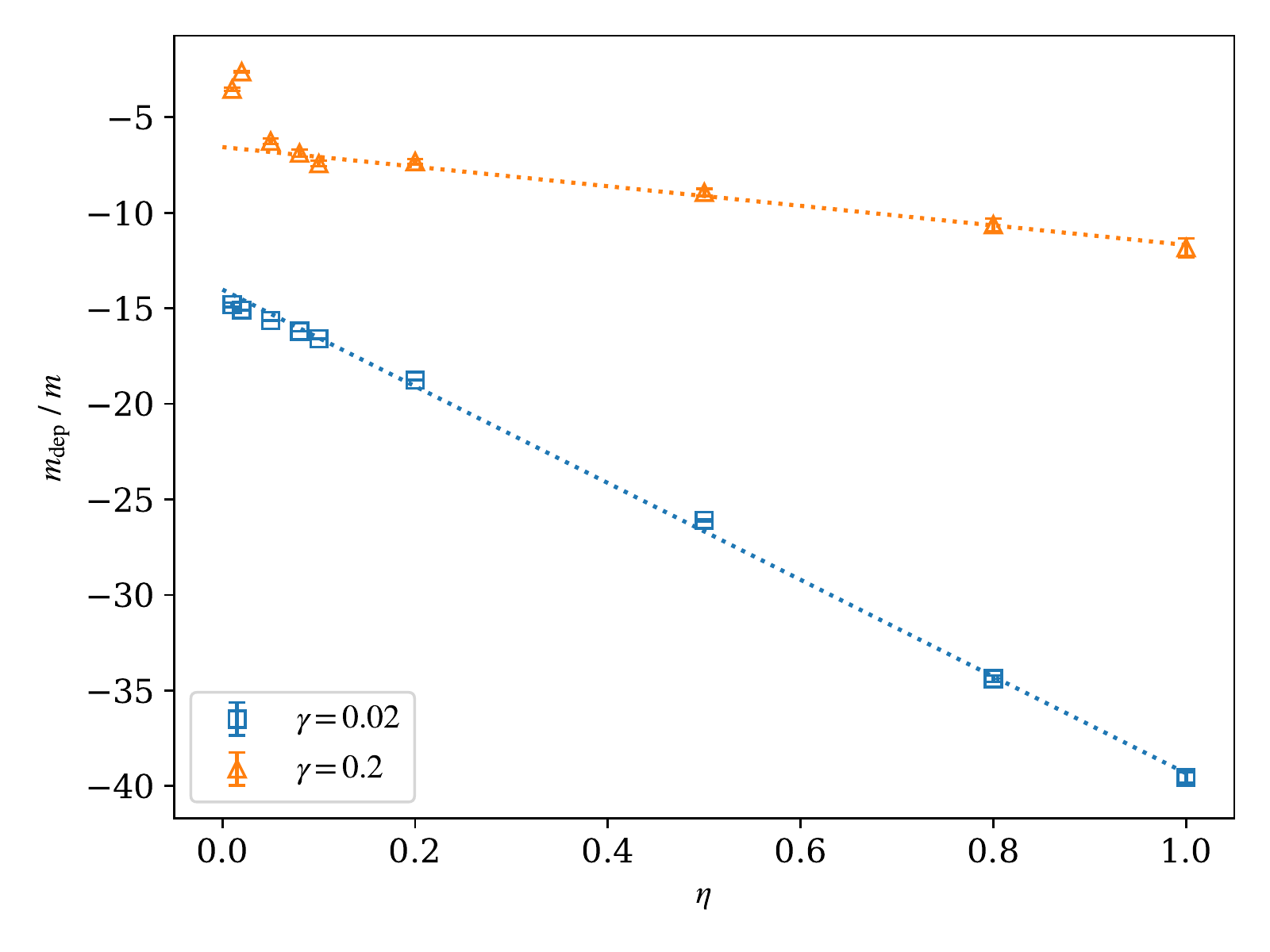} & \includegraphics[height=6.5 cm]{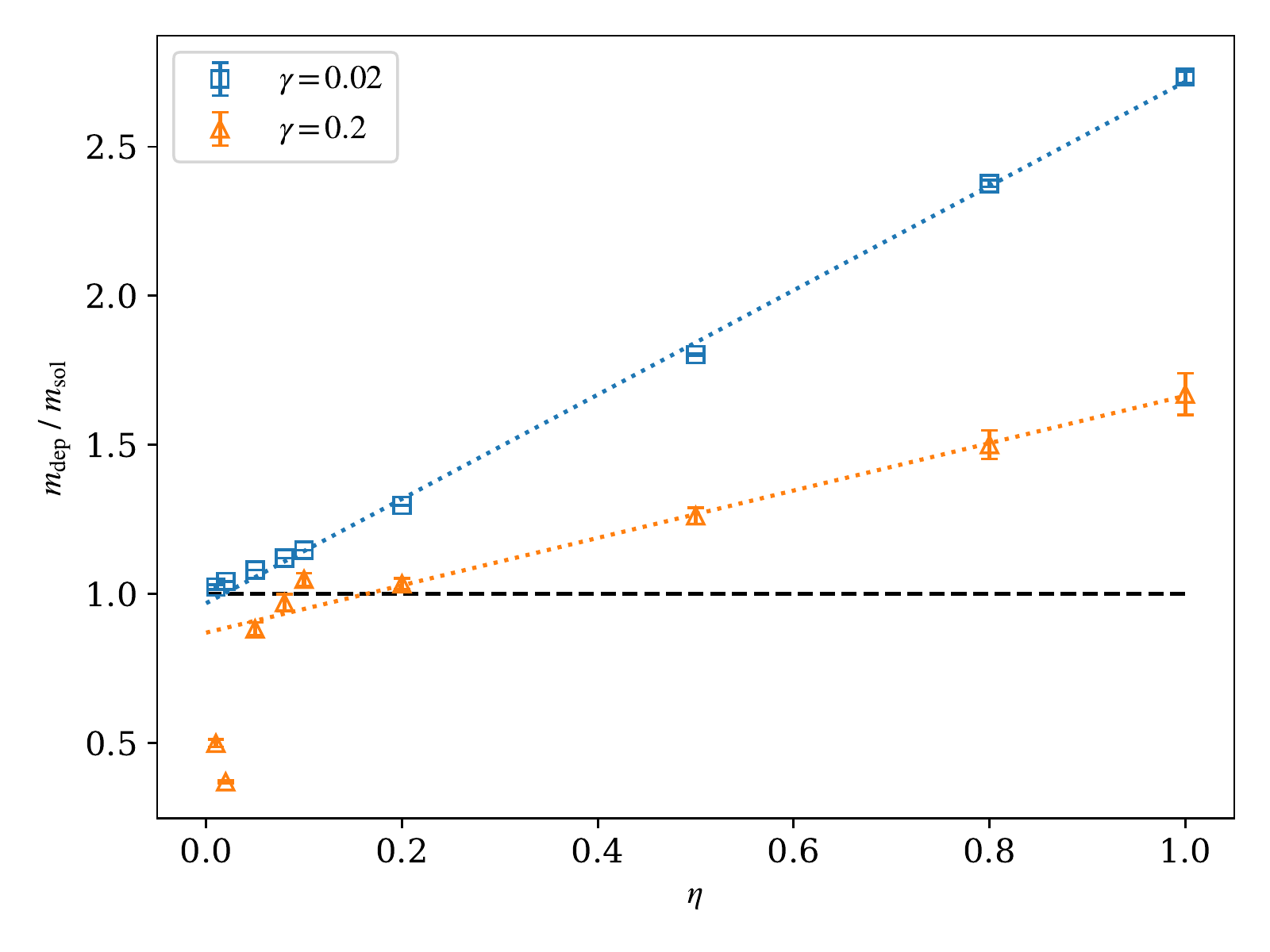} 
    \end{tabular}
    \caption{\label{fig:depleton-mass}  Effective mass at  half-umklapp.
    (a) Effective mass of the Bose gas with impurity at the half-umklapp point $P=N_\mathrm{tot} P_0/2$ (depleton mass $m_\mathrm{dep}$) in units of the bare mass $m$ as a function of the impurity coupling strength $\eta$. The dotted lines are linear fits to the data with $\eta>\gamma$ and highlight the linear trends.
    (b) Ratio of the effective mass of the impurity (depleton mass $m_\mathrm{dep}$)  to the effective mass of the pure Bose gas (soliton mass $m_\mathrm{sol}$)
at the half-umklapp point $P=N_\mathrm{tot} P_0/2$.
 The dashed line in (b) indicates a depleton/soliton mass ratio is 1. The soliton mass is $m_\mathrm{sol} = -14.47232(5)m$ for $\gamma=0.02$ and $m_\mathrm{sol}=-7.079(34)m$ for $\gamma=0.2$.}
\end{figure}
\begin{paracol}{2}
\switchcolumn

Figure~\ref{fig:depleton-mass} shows the extracted effective mass as a function of the impurity coupling strength $\eta$ for two different values of the Bose gas interaction constant $\gamma$. In the regime $\eta>\gamma$ our data shows a linear trend with $\eta$. A linear dependence of the effective mass on $\eta$ is expected from exact results for an equal-mass impurity in a Tonks–Girardeau gas ($\gamma=\infty$) \cite{Schecter2016}.
The effective mass becomes particularly heavy for small $\gamma$ and large $\eta$, up to several times the mass of the dark soliton at the same value of $\gamma$, as seen in Fig.~\ref{fig:depleton-mass}(b). As the magnitude of the extracted effective mass (from the finite-size corrected dispersion relation) becomes larger than the total system mass of $20m$, we call this the super-heavy regime. Note that without  the finite-size correction of Eq.~\eqref{eq:Omega}, the curvature of the yrast dispersion changes from concave to convex, which means that the uncorrected effective mass diverges and changes sign (not shown).
The heavy effective mass regime has potential experimental relevance, as it is relevant for realizing  physical phenomena such as  Bloch oscillations \cite{Schecter2012,Schecter2016}. Furthermore, a recent study demonstrates that the dynamical phenomenon of temporal orthogonality catastrophe is exhibited, given the impurity-boson couplings are sufficiently stronger than the intra-species background ones \cite{Koutentakis2021}.

Another interesting feature shown in Fig.~\ref{fig:depleton-mass}(b) is that the impurity effective mass is approximately equal to the soliton mass for $\eta=\gamma$. In the regime where $\eta < \gamma$ (seen for $\gamma=0.2$), the effective mass becomes very small in magnitude, trending towards zero. This is consistent with the establishment of a cusp in the dispersion relation at the half-umklapp point, a feature that was already discussed in Sec.~\ref{sec:yrast_disp}. While the concept of an effective mass breaks down in the cusp regime, our data can be used to determine that the transition happens approximately where $\eta=\gamma$, thus shedding some light on the question of the critical coupling which remains unsolved from Ref~\cite{Lamacraft2009}.

\subsection{Spin-flip Energy} \label{sec:spin-flip}

So far we have considered the yrast excitation energies, which measure how much energy is required to deposit momentum into the system on top of the energy of the ground state at $P=0$. Now we want to examine the energy that is required to flip a spin in the Bose gas at fixed momentum. We define the spin-flip energy $E_\mathrm{SF}(P)$ as
\begin{align}
E_\mathrm{SF}(P) = E_{N_\mathrm{tot}-1,1}(P) - E_{N_\mathrm{tot},0}(P) .
\end{align}

\end{paracol}
\begin{figure}[htb]
    \widefigure
    \begin{tabular}{cc}
        (a) & (b) \\
        \includegraphics[height=6.5 cm]{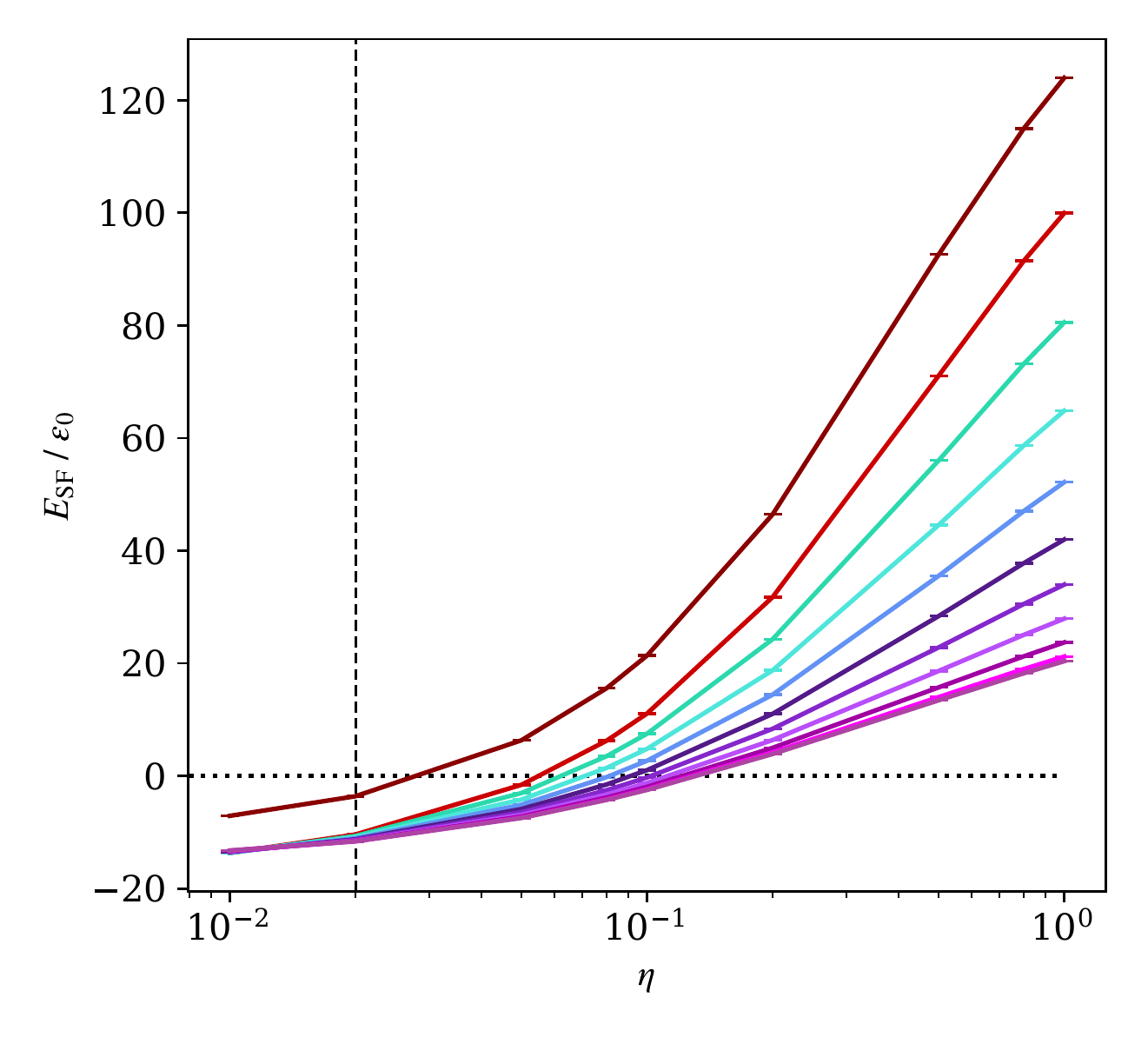} & \includegraphics[height=6.5 cm]{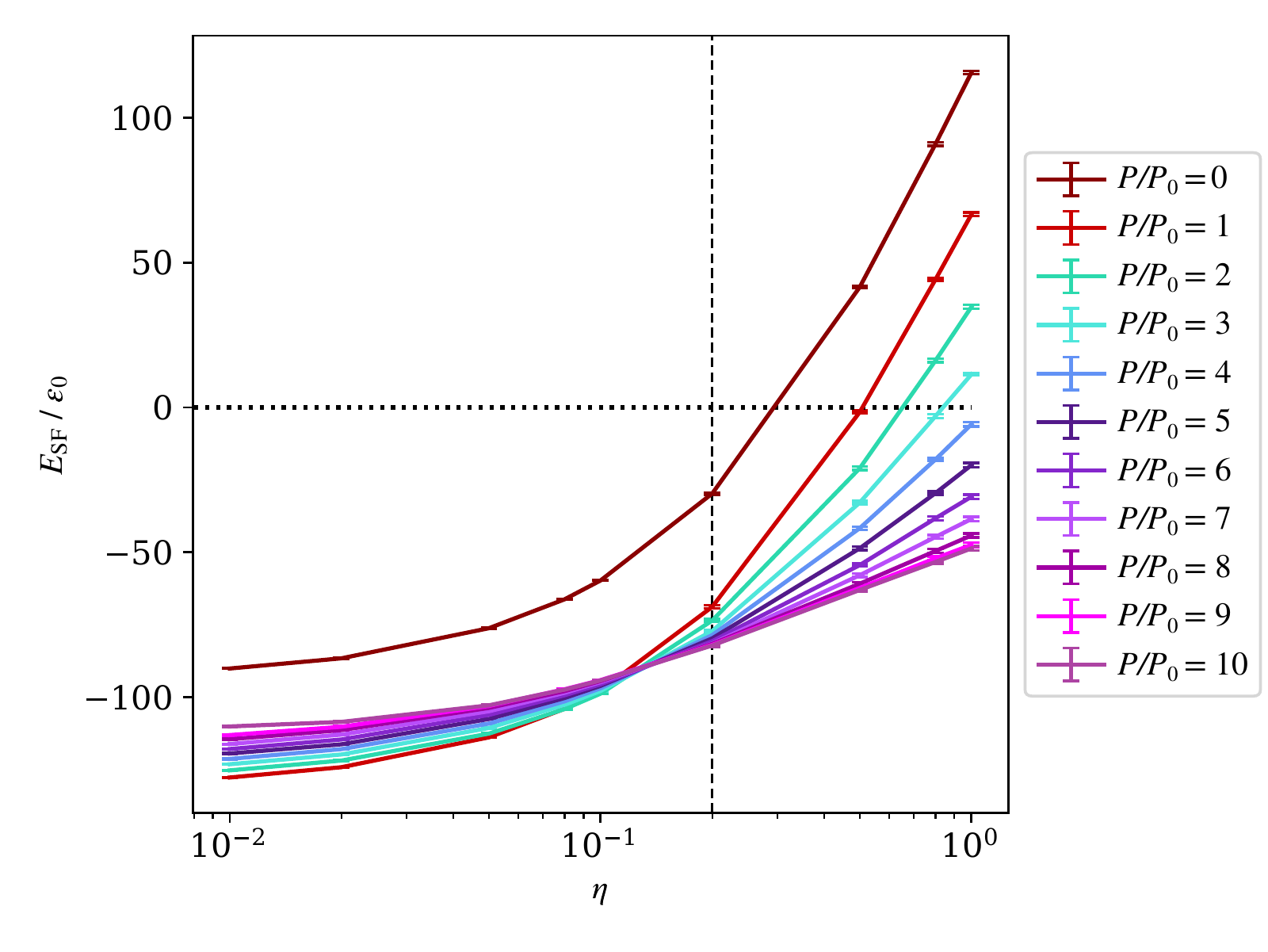}
    \end{tabular}
    \caption{\label{fig:E-SF}The spin-flip energy as a function of the boson-impurity coupling strength $\eta$, with boson-boson coupling strength (a) $\gamma=0.02$ and (b) $\gamma=0.2$.The data for $P>10$ are not presented, as they follow the symmetry across $P=10$ in the yrast spectrum, hence are overlapping with existing data on this figure.
    }
\end{figure}
\begin{paracol}{2}
\switchcolumn

Figure~\ref{fig:E-SF} shows the spin-flip energy as a function of the impurity coupling strength $\eta$ for yrast states at different total momentum $P$. The two panels refer to different values of the boson-boson interaction strength $\gamma$. 
The brown data shows the spin-flip energy for the $P=0$ ground state. It is separated by a gap from the spin-flip energies at other momentum values, which are all lower. The ground state spin-flip energy increases with $\eta$ and crosses zero, meaning that for large $\eta$ flipping the spin becomes energetically unfavorable. 
The vertical lines indicate where $\eta=\gamma$. At this point the impurity is distinguishable from the background Bose gas but all physical properties such as mass and interactions are the same. The spin-flip energy is thus solely due to quantum statistics. 
From the data shown in Fig.~\ref{fig:E-SF} we see that the spin-flip energies are all negative at this point, and thus the system with impurity has lower energy than the pure Bose gas, i.e.\ it is favorable to flip the spin, for any value of the total momentum $P$. Furthermore, the energy gain is larger the higher the momentum (up to the half umklapp value). This trend is remarkably not maintained when $\eta<\gamma$ as seen in Fig.~\ref{fig:E-SF}(b). This behavior can be rationalized from the quantitative changes in the yrast dispersions  shown in Fig.~\ref{fig:yrast-gamma}.

\section{Conclusions}\label{sec:conclusions}

Using the FCIQMC method, we investigated the properties of the yrast states of Bose gases coupled with a mobile impurity in one spatial dimension. Based on the energies and the first and second order correlation functions of yrast states, we identified the polaron and depleton regimes, as well as the transitions between them. The extracted depleton effective mass revealed a super-heavy regime where the magnitude of the (negative) depleton mass exceeds the mass of the finite Bose gas. We also observed a qualitative change in behavior crossing $\eta=\gamma$ in all calculated quantities. For the $\eta>\gamma$ regime we can identify the formation of depletons around the half-umklapp point where the impurity is more or less confined to the density hole of the gray/dark soliton of the Bose gas. The depleton picture becomes inadequate for smaller interactions between the impurity and the bosons, $\eta<\gamma$, with the impurity not longer hybridizing with the soliton. This behavior is consistent with an observed break-down of the effective mass concept below $\eta=\gamma$.

In this work, the FCIQMC method is applied to a bosonic many-body problem for the first time. Due to the non-stoquastic nature of the momentum-space Hamiltonian~\eqref{eq:2c-ham-mom}, the sign problem exists and becomes severe when either $\eta$ or $\gamma$ is large. Through this study, we demonstrated the effective suppression of the sign problem in FCIQMC by the application of the initiator approximation, showing the potential of FCIQMC for studying complex bosonic many-body systems.

\textit{Outlook:} The demonstrated computational method is extremely versatile and can be applied to a wide range of physics question. Possible future extensions of this study include extrapolating the results to the thermodynamic limit. A transcorrelated Hamiltonian \cite{Jeszenszki2018,Jeszenszki2020} can be applied to accelerate the basis set convergence to the infinite limit. There are also many interesting set-ups that we wish to study further.
In this work, we only focus on the cases where the impurity and bosons all have identical mass and repulsive interactions, which could be extended to unequal masses and attractively interacting impurities. Attractive impurity-boson coupling has been studied in the polaron regime \cite{Parisi2017,Panochko2019} but not yet explored in the context of depleton physics. In addition, the case of an impurity in a strongly interacting Bose gas {or with long-range interactions} is interesting to study, where perturbative and mean-field approaches are {of limited use or} invalid. A more complex system with two impurity atoms, known as the bipolaron problem {\cite{Keiler2020,Dutta2013,Camacho-Guardian2018,Will2021,Petkovic2021}}, at non-zero momentum is also interesting due to its connection to high-temperature superconductivity {\cite{Camacho-Guardian2018,Will2021}}.

\vspace{6pt} 



\authorcontributions{Conceptualization, J.B. and E.P.; data curation, M.Y. and M.\v{C}.; writing---original draft preparation, M.Y. and M.\v{C}.; visualization, M.Y.; supervision, J.B and E.P.; project administration, J.B.; funding acquisition, J.B.; M.\v{C}. made special contribution to the software optimization and testing.
All authors contributed to the methodology, software development, data analysis, and review and editing of the manuscript. 
All authors have read and agreed to the published version of the manuscript.}


\funding{This research was funded by Marsden Fund of New Zealand, Contract No.\ MAU1604, from government funding managed by the Royal Society of New Zealand Te Apārangi.}

\dataavailability{The data that support the findings of this study are obtainable with \texttt{Rimu.jl}. The \texttt{Rimu.jl} program  library  is available as an open source project on GitHub. The code can be obtained at \url{https://github.com/joachimbrand/Rimu.jl}.}


\acknowledgments{The authors wish to acknowledge the use of New Zealand eScience Infrastructure (NeSI) high performance computing facilities and consulting support as part of this research. New Zealand's national facilities are provided by NeSI and funded jointly by NeSI's collaborator institutions and through the Ministry of Business, Innovation \& Employment's Research Infrastructure programme. URL \url{https://www.nesi.org.nz}. The authors further acknowledge the use of Massey University's CTCP high-performance cluster, and Mike Yap for technical support.}


\conflictsofinterest{The authors declare no conflict of interest.} 





\appendixtitles{yes} 
\appendixstart
\appendix
\section{Eliminating Biases} \label{sec:Bias}
While bosonic systems can often be described by stoquastic Hamiltonians characterized by having only non-positive, real off-diagonal elements, 
the momentum-space Hamiltonian of Eq.~\ref{eq:2c-ham-mom} considered here, is non-stoquastic. As a consequence, one has to deal with the QMC sign problem, that originates from the fact that different configurations can spawn into the same configuration with incoherent signs.   

\begin{figure}[htb]
    \centering
    \includegraphics[height=6.5 cm]{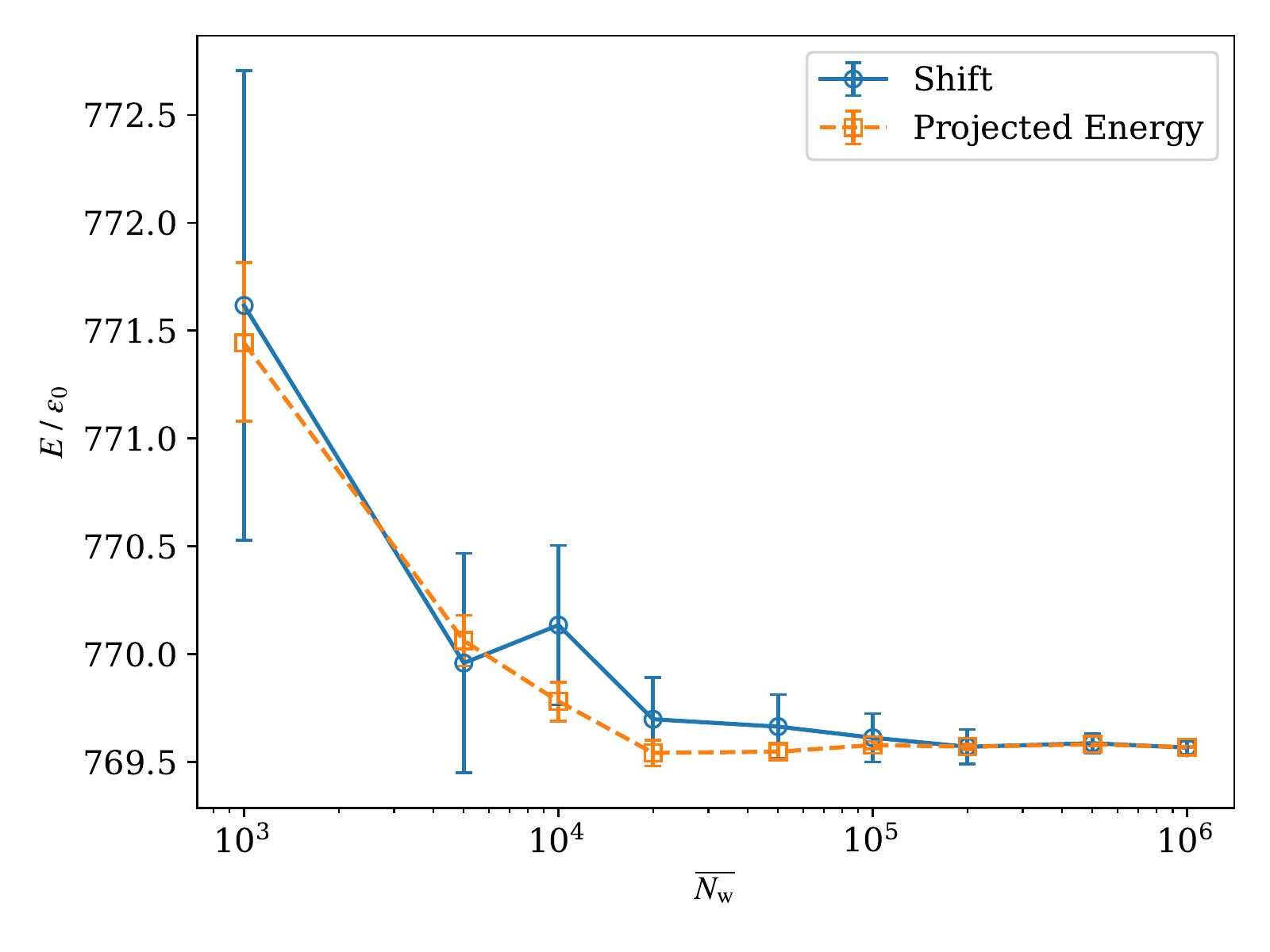}
    \caption{\label{fig:bias} Ground-state energy against the equilibrated walker numbers, $\overline{N_\mathrm{w}}$. The system size is the same as used in other sections. The boson-boson coupling is $\gamma=0.2$ and the impurity-boson coupling is $\eta=2$, This is a stronger interaction  than in our other calculations and should lead to the largest bias. For $\overline{N_\mathrm{w}}>10^4$ the bias becomes smaller than the statistical errors. }
\end{figure}

In FCIQMC, the initiator approach can be used to mitigate this sign problem by restricting the walker spawning process to the dominant configurations. This enforces a better coherence in the sign structure of the wave function. Albeit typically small \cite{Cleland2010,Booth2014}, an initiator bias can be observed as a consequence of the initiator approximation when an insufficient number of walkers is used to sample a much larger Hilbert space. Furthermore, the population control bias \cite{Umrigar1993,Brand2021,Ghanem2021} is a  stochastic bias that appears as a result of sampling noise. It is typically much smaller than the initiator bias (where the latter is applicable) and scales like a power law with the number of walkers \cite{Brand2021}. Both biases can be reduced below the size of statistical error bars by increasing the walker population.  To make sure our calculated energies are bias free and a sufficiently large walker population is used, one can check the ground-state energy as a function of the equilibrated walker number, $\overline{N_\mathrm{w}}$, as shown in Fig.~\ref{fig:bias}.

It can be seen that when $\overline{N_\mathrm{w}}>10^4$ the biases in the shift and projected energies are smaller than the statistical errors, and converged to the same energy throughout. This convergence check is carried out with a larger interaction strength ($\eta=2, \gamma=0.2$) than used for any of the data presented in the main article, and thus should overestimate the bias for the presented data.  
For all energies presented in Sec.~\ref{sec:results}, a walker population of $\overline{N_\mathrm{w}}=10^6$ is used. Hence we are confident that the presented data is  free of both, the initiator and the population control bias.


\end{paracol}
\reftitle{References}


\externalbibliography{yes}
\bibliography{bib-ray,fciqmc,bib-further}

\begin{thebibliography}{999}

\bibitem[Emin(2012)]{Emin2012}
Emin, D.
\newblock {\em {Polarons}}; Cambridge University Press: Cambridge,  2012.
\newblock
  doi:{\changeurlcolor{black}\href{https://doi.org/10.1017/CBO9781139023436}{\detokenize{10.1017/CBO9781139023436}}}.

\bibitem[Alexandrov and Devreese(2010)]{Alexandrov2010}
Alexandrov, A.S.; Devreese, J.T.
\newblock {\em {Advances in Polaron Physics}}; Vol. 159, {\em Springer Series
  in Solid-State Sciences}, Springer Berlin Heidelberg: Berlin, Heidelberg,
  2010.
\newblock
  doi:{\changeurlcolor{black}\href{https://doi.org/10.1007/978-3-642-01896-1}{\detokenize{10.1007/978-3-642-01896-1}}}.

\bibitem[Landau and Pekar(1948)]{Landau1948a}
Landau, L.D.; Pekar, S.I.
\newblock {Effective mass of a polaron}.
\newblock {\em Zh. Eksp. Teor. Fiz} {\bf 1948}, {\em 18},~419--423.

\bibitem[Bardeen \em{et~al.}(1966)Bardeen, Baym, and Pines]{Bardeen1966}
Bardeen, J.; Baym, G.; Pines, D.
\newblock {Interactions Between He3 Atoms in Dilute Solutions of He3 in
  Superfluid He4}.
\newblock {\em Phys. Rev. Lett.} {\bf 1966}, {\em 17},~372--375.
\newblock
  doi:{\changeurlcolor{black}\href{https://doi.org/10.1103/PhysRevLett.17.372}{\detokenize{10.1103/PhysRevLett.17.372}}}.

\bibitem[Chin \em{et~al.}(2010)Chin, Grimm, Julienne, and Tiesinga]{Chin2010}
Chin, C.; Grimm, R.; Julienne, P.; Tiesinga, E.
\newblock {Feshbach resonances in ultracold gases}.
\newblock {\em Rev. Mod. Phys.} {\bf 2010}, {\em 82},~1225--1286,
  \href{http://xxx.lanl.gov/abs/0812.1496}{{\normalfont [0812.1496]}}.
\newblock
  doi:{\changeurlcolor{black}\href{https://doi.org/10.1103/RevModPhys.82.1225}{\detokenize{10.1103/RevModPhys.82.1225}}}.

\bibitem[Vale and Zwierlein(2021)]{Vale2021a}
Vale, C.J.; Zwierlein, M.
\newblock {Spectroscopic probes of quantum gases}.
\newblock {\em Nat. Phys.} {\bf 2021}, {\em 17},~1305--1315.
\newblock
  doi:{\changeurlcolor{black}\href{https://doi.org/10.1038/s41567-021-01434-6}{\detokenize{10.1038/s41567-021-01434-6}}}.

\bibitem[J{\o}rgensen \em{et~al.}(2016)J{\o}rgensen, Wacker, Skalmstang,
  Parish, Levinsen, Christensen, Bruun, and Arlt]{Jorgensen2016}
J{\o}rgensen, N.B.; Wacker, L.; Skalmstang, K.T.; Parish, M.M.; Levinsen, J.;
  Christensen, R.S.; Bruun, G.M.; Arlt, J.J.
\newblock {Observation of Attractive and Repulsive Polarons in a Bose-Einstein
  Condensate}.
\newblock {\em Phys. Rev. Lett.} {\bf 2016}, {\em 117},~1--6,
  \href{http://xxx.lanl.gov/abs/1604.07883}{{\normalfont [1604.07883]}}.
\newblock
  doi:{\changeurlcolor{black}\href{https://doi.org/10.1103/PhysRevLett.117.055302}{\detokenize{10.1103/PhysRevLett.117.055302}}}.

\bibitem[Hu \em{et~al.}(2016)Hu, {Van De Graaff}, Kedar, Corson, Cornell, and
  Jin]{Hu2016}
Hu, M.G.; {Van De Graaff}, M.J.; Kedar, D.; Corson, J.P.; Cornell, E.A.; Jin,
  D.S.
\newblock {Bose Polarons in the Strongly Interacting Regime}.
\newblock {\em Phys. Rev. Lett.} {\bf 2016}, {\em 117},~1--6,
  \href{http://xxx.lanl.gov/abs/1605.00729}{{\normalfont [1605.00729]}}.
\newblock
  doi:{\changeurlcolor{black}\href{https://doi.org/10.1103/PhysRevLett.117.055301}{\detokenize{10.1103/PhysRevLett.117.055301}}}.

\bibitem[Yan \em{et~al.}(2020)Yan, Ni, Robens, and Zwierlein]{Yan2020}
Yan, Z.Z.; Ni, Y.; Robens, C.; Zwierlein, M.W.
\newblock {Bose polarons near quantum criticality}.
\newblock {\em Science (80-. ).} {\bf 2020}, {\em 368},~190--194,
  \href{http://xxx.lanl.gov/abs/1904.02685}{{\normalfont [1904.02685]}}.
\newblock
  doi:{\changeurlcolor{black}\href{https://doi.org/10.1126/science.aax5850}{\detokenize{10.1126/science.aax5850}}}.

\bibitem[Skou \em{et~al.}(2021)Skou, Skov, J{\o}rgensen, Nielsen,
  Camacho-Guardian, Pohl, Bruun, and Arlt]{Skou2021}
Skou, M.G.; Skov, T.G.; J{\o}rgensen, N.B.; Nielsen, K.K.; Camacho-Guardian,
  A.; Pohl, T.; Bruun, G.M.; Arlt, J.J.
\newblock {Non-equilibrium quantum dynamics and formation of the Bose polaron}.
\newblock {\em Nat. Phys.} {\bf 2021}, {\em 17},~731--735,
  \href{http://xxx.lanl.gov/abs/2005.00424}{{\normalfont [2005.00424]}}.
\newblock
  doi:{\changeurlcolor{black}\href{https://doi.org/10.1038/s41567-021-01184-5}{\detokenize{10.1038/s41567-021-01184-5}}}.

\bibitem[Imambekov \em{et~al.}(2012)Imambekov, Schmidt, and
  Glazman]{Imambekov2012}
Imambekov, A.; Schmidt, T.L.; Glazman, L.I.
\newblock {One-dimensional quantum liquids: Beyond the Luttinger liquid
  paradigm}.
\newblock {\em Rev. Mod. Phys.} {\bf 2012}, {\em 84},~1253--1306.
\newblock
  doi:{\changeurlcolor{black}\href{https://doi.org/10.1103/RevModPhys.84.1253}{\detokenize{10.1103/RevModPhys.84.1253}}}.

\bibitem[Cherny \em{et~al.}(2012)Cherny, Caux, and Brand]{Cherny2012}
Cherny, A.Y.; Caux, J.S.; Brand, J.
\newblock {Theory of superfluidity and drag force in the one-dimensional Bose
  gas}.
\newblock {\em Front. Phys.} {\bf 2012}, {\em 7},~54--71,
  \href{http://xxx.lanl.gov/abs/1106.6329}{{\normalfont [1106.6329]}}.
\newblock
  doi:{\changeurlcolor{black}\href{https://doi.org/10.1007/s11467-011-0211-2}{\detokenize{10.1007/s11467-011-0211-2}}}.

\bibitem[Gangardt and Kamenev(2009)]{Gangardt2009}
Gangardt, D.M.; Kamenev, A.
\newblock {Bloch oscillations in a one-dimensional spinor gas}.
\newblock {\em Phys. Rev. Lett.} {\bf 2009}, {\em 102}.
\newblock
  doi:{\changeurlcolor{black}\href{https://doi.org/10.1103/PhysRevLett.102.070402}{\detokenize{10.1103/PhysRevLett.102.070402}}}.

\bibitem[Schecter \em{et~al.}(2012)Schecter, Gangardt, and
  Kamenev]{Schecter2012}
Schecter, M.; Gangardt, D.M.; Kamenev, A.
\newblock {Dynamics and Bloch oscillations of mobile impurities in
  one-dimensional quantum liquids}.
\newblock {\em Ann. Phys. (N. Y).} {\bf 2012}, {\em 327},~639--670,
  \href{http://xxx.lanl.gov/abs/1105.6136}{{\normalfont [1105.6136]}}.
\newblock
  doi:{\changeurlcolor{black}\href{https://doi.org/10.1016/j.aop.2011.10.001}{\detokenize{10.1016/j.aop.2011.10.001}}}.

\bibitem[Bloch(1929)]{Bloch1929}
Bloch, F.
\newblock {\"Uber die Quantenmechanik der Elektronen in Kristallgittern}.
\newblock {\em Z. Phys.} {\bf 1929}, {\em 52},~555--600.
\newblock
  doi:{\changeurlcolor{black}\href{https://doi.org/10.1007/BF01339455}{\detokenize{10.1007/BF01339455}}}.

\bibitem[Feldmann \em{et~al.}(1992)Feldmann, Leo, Shah, Miller, Cunningham,
  Meier, von Plessen, Schulze, Thomas, and Schmitt-Rink]{Feldmann1992}
Feldmann, J.; Leo, K.; Shah, J.; Miller, D.A.B.; Cunningham, J.E.; Meier, T.;
  von Plessen, G.; Schulze, A.; Thomas, P.; Schmitt-Rink, S.
\newblock {Optical investigation of Bloch oscillations in a semiconductor
  superlattice}.
\newblock {\em Phys. Rev. B} {\bf 1992}, {\em 46},~7252--7255.
\newblock
  doi:{\changeurlcolor{black}\href{https://doi.org/10.1103/PhysRevB.46.7252}{\detokenize{10.1103/PhysRevB.46.7252}}}.

\bibitem[{Ben Dahan} \em{et~al.}(1996){Ben Dahan}, Peik, Reichel, Castin, and
  Salomon]{BenDahan1996}
{Ben Dahan}, M.; Peik, E.; Reichel, J.; Castin, Y.; Salomon, C.
\newblock {Bloch Oscillations of Atoms in an Optical Potential}.
\newblock {\em Phys. Rev. Lett.} {\bf 1996}, {\em 76},~4508--4511.
\newblock
  doi:{\changeurlcolor{black}\href{https://doi.org/10.1103/PhysRevLett.76.4508}{\detokenize{10.1103/PhysRevLett.76.4508}}}.

\bibitem[Gamayun \em{et~al.}(2014)Gamayun, Lychkovskiy, and
  Cheianov]{Gamayun2014}
Gamayun, O.; Lychkovskiy, O.; Cheianov, V.
\newblock {Kinetic theory for a mobile impurity in a degenerate Tonks-Girardeau
  gas}.
\newblock {\em Phys. Rev. E} {\bf 2014}, {\em 90},~32132,
  \href{http://xxx.lanl.gov/abs/1402.6362}{{\normalfont [1402.6362]}}.
\newblock
  doi:{\changeurlcolor{black}\href{https://doi.org/10.1103/PhysRevE.90.032132}{\detokenize{10.1103/PhysRevE.90.032132}}}.

\bibitem[Schecter \em{et~al.}(2016)Schecter, Gangardt, and
  Kamenev]{Schecter2016}
Schecter, M.; Gangardt, D.M.; Kamenev, A.
\newblock {Quantum impurities: From mobile Josephson junctions to depletons}.
\newblock {\em New J. Phys.} {\bf 2016}, {\em 18},~65002,
  \href{http://xxx.lanl.gov/abs/1601.00628}{{\normalfont [1601.00628]}}.
\newblock
  doi:{\changeurlcolor{black}\href{https://doi.org/10.1088/1367-2630/18/6/065002}{\detokenize{10.1088/1367-2630/18/6/065002}}}.

\bibitem[Meinert \em{et~al.}(2017)Meinert, Knap, Kirilov, Jag-Lauber, Zvonarev,
  Demler, and N{\"{a}}gerl]{Meinert2017}
Meinert, F.; Knap, M.; Kirilov, E.; Jag-Lauber, K.; Zvonarev, M.B.; Demler, E.;
  N{\"{a}}gerl, H.C.
\newblock {Bloch oscillations in the absence of a lattice}.
\newblock {\em Science} {\bf 2017}, {\em 356},~945--948,
  \href{http://xxx.lanl.gov/abs/1608.08200}{{\normalfont [1608.08200]}}.
\newblock
  doi:{\changeurlcolor{black}\href{https://doi.org/10.1126/science.aah6616}{\detokenize{10.1126/science.aah6616}}}.

\bibitem[Palzer \em{et~al.}(2009)Palzer, Zipkes, Sias, and
  K{\"{o}}hl]{palzer09}
Palzer, S.; Zipkes, C.; Sias, C.; K{\"{o}}hl, M.
\newblock {Quantum Transport through a Tonks-Girardeau Gas}.
\newblock {\em Phys. Rev. Lett.} {\bf 2009}, {\em 103},~150601.
\newblock
  doi:{\changeurlcolor{black}\href{https://doi.org/10.1103/PhysRevLett.103.150601}{\detokenize{10.1103/PhysRevLett.103.150601}}}.

\bibitem[Fukuhara \em{et~al.}(2013)Fukuhara, Kantian, Endres, Cheneau,
  Schau{\ss}, Hild, Bellem, Schollw{\"{o}}ck, Giamarchi, Gross, Bloch, and
  Kuhr]{Fukuhara2013a}
Fukuhara, T.; Kantian, A.; Endres, M.; Cheneau, M.; Schau{\ss}, P.; Hild, S.;
  Bellem, D.; Schollw{\"{o}}ck, U.; Giamarchi, T.; Gross, C.; Bloch, I.; Kuhr,
  S.
\newblock {Quantum dynamics of a mobile spin impurity}.
\newblock {\em Nat. Phys.} {\bf 2013}, {\em 9},~235--241.
\newblock
  doi:{\changeurlcolor{black}\href{https://doi.org/10.1038/nphys2561}{\detokenize{10.1038/nphys2561}}}.

\bibitem[Catani \em{et~al.}(2012)Catani, Lamporesi, Naik, Gring, Inguscio,
  Minardi, Kantian, and Giamarchi]{Catani2012}
Catani, J.; Lamporesi, G.; Naik, D.; Gring, M.; Inguscio, M.; Minardi, F.;
  Kantian, A.; Giamarchi, T.
\newblock {Quantum dynamics of impurities in a one-dimensional Bose gas}.
\newblock {\em Phys. Rev. A} {\bf 2012}, {\em 85},~023623.
\newblock
  doi:{\changeurlcolor{black}\href{https://doi.org/10.1103/PhysRevA.85.023623}{\detokenize{10.1103/PhysRevA.85.023623}}}.

\bibitem[Spethmann \em{et~al.}(2012)Spethmann, Kindermann, John, Weber,
  Meschede, and Widera]{Spethmann2012}
Spethmann, N.; Kindermann, F.; John, S.; Weber, C.; Meschede, D.; Widera, A.
\newblock {Dynamics of Single Neutral Impurity Atoms Immersed in an Ultracold
  Gas}.
\newblock {\em Phys. Rev. Lett.} {\bf 2012}, {\em 109},~235301,
  \href{http://xxx.lanl.gov/abs/1204.6051}{{\normalfont [1204.6051]}}.
\newblock
  doi:{\changeurlcolor{black}\href{https://doi.org/10.1103/PhysRevLett.109.235301}{\detokenize{10.1103/PhysRevLett.109.235301}}}.

\bibitem[Kain and Ling(2018)]{Kain2018}
Kain, B.; Ling, H.Y.
\newblock {Analytical study of static beyond-Fr{\"{o}}hlich Bose polarons in
  one dimension}.
\newblock {\em Phys. Rev. A} {\bf 2018}, {\em 98},
  \href{http://xxx.lanl.gov/abs/1809.10601}{{\normalfont [1809.10601]}}.
\newblock
  doi:{\changeurlcolor{black}\href{https://doi.org/10.1103/PhysRevA.98.033610}{\detokenize{10.1103/PhysRevA.98.033610}}}.

\bibitem[Panochko and Pastukhov(2019)]{Panochko2019}
Panochko, G.; Pastukhov, V.
\newblock {Mean-field construction for spectrum of one-dimensional Bose
  polaron}.
\newblock {\em Ann. Phys. (N. Y).} {\bf 2019}, {\em 409},~167933,
  \href{http://xxx.lanl.gov/abs/1903.05953}{{\normalfont [1903.05953]}}.
\newblock
  doi:{\changeurlcolor{black}\href{https://doi.org/10.1016/j.aop.2019.167933}{\detokenize{10.1016/j.aop.2019.167933}}}.

\bibitem[Dutta and Mueller(2013)]{Dutta2013}
Dutta, S.; Mueller, E.J.
\newblock {Variational study of polarons and bipolarons in a one-dimensional
  Bose lattice gas in both the superfluid and the Mott-insulator regimes}.
\newblock {\em Phys. Rev. A} {\bf 2013}, {\em 88}.
\newblock
  doi:{\changeurlcolor{black}\href{https://doi.org/10.1103/PhysRevA.88.053601}{\detokenize{10.1103/PhysRevA.88.053601}}}.

\bibitem[Koutentakis \em{et~al.}(2021)Koutentakis, Mistakidis, and
  Schmelcher]{Koutentakis2021}
Koutentakis, G.M.; Mistakidis, S.I.; Schmelcher, P.
\newblock {Pattern formation in one-dimensional polaron systems and temporal
  orthogonality catastrophe} {\bf 2021}.
\newblock  \href{http://xxx.lanl.gov/abs/2110.11165}{{\normalfont
  [2110.11165]}}.

\bibitem[Seetharam \em{et~al.}(2021)Seetharam, Shchadilova, Grusdt, Zvonarev,
  and Demler]{Seetharam}
Seetharam, K.; Shchadilova, Y.; Grusdt, F.; Zvonarev, M.; Demler, E.
\newblock {Quantum Cherenkov transition of finite momentum Bose polarons} {\bf
  2021}.
\newblock  \href{http://xxx.lanl.gov/abs/2109.12260}{{\normalfont
  [2109.12260]}}.

\bibitem[Ichmoukhamedov and Tempere(2019)]{Ichmoukhamedov2019}
Ichmoukhamedov, T.; Tempere, J.
\newblock {Feynman path-integral treatment of the Bose polaron beyond the
  Fr{\"{o}}hlich model}.
\newblock {\em Phys. Rev. A} {\bf 2019}, {\em 100},~43605,
  \href{http://xxx.lanl.gov/abs/1905.07368}{{\normalfont [1905.07368]}}.
\newblock
  doi:{\changeurlcolor{black}\href{https://doi.org/10.1103/PhysRevA.100.043605}{\detokenize{10.1103/PhysRevA.100.043605}}}.

\bibitem[Jager and Barnett(2021)]{Jager2021a}
Jager, J.; Barnett, R.
\newblock {Stochastic-field approach to the quench dynamics of the
  one-dimensional Bose polaron}.
\newblock {\em Phys. Rev. Res.} {\bf 2021}, {\em 3},~033212,
  \href{http://xxx.lanl.gov/abs/2103.13457}{{\normalfont [2103.13457]}}.
\newblock
  doi:{\changeurlcolor{black}\href{https://doi.org/10.1103/PhysRevResearch.3.033212}{\detokenize{10.1103/PhysRevResearch.3.033212}}}.

\bibitem[Volosniev and Hammer(2017)]{Volosniev2017}
Volosniev, A.G.; Hammer, H.W.
\newblock {Analytical approach to the Bose-polaron problem in one dimension}.
\newblock {\em Phys. Rev. A} {\bf 2017}, {\em 96},~31601,
  \href{http://xxx.lanl.gov/abs/1704.00622}{{\normalfont [1704.00622]}}.
\newblock
  doi:{\changeurlcolor{black}\href{https://doi.org/10.1103/PhysRevA.96.031601}{\detokenize{10.1103/PhysRevA.96.031601}}}.

\bibitem[Grusdt \em{et~al.}(2015)Grusdt, Shchadilova, Rubtsov, and
  Demler]{Grusdt2015}
Grusdt, F.; Shchadilova, Y.E.; Rubtsov, A.N.; Demler, E.
\newblock {Renormalization group approach to the Fr{\"{o}}hlich polaron model:
  Application to impurity-BEC problem}.
\newblock {\em Sci. Rep.} {\bf 2015}, {\em 5},~1--14,
  \href{http://xxx.lanl.gov/abs/1410.2203}{{\normalfont [1410.2203]}}.
\newblock
  doi:{\changeurlcolor{black}\href{https://doi.org/10.1038/srep12124}{\detokenize{10.1038/srep12124}}}.

\bibitem[Isaule \em{et~al.}(2021)Isaule, Morera, Massignan, and
  Juli{\'{a}}-D{\'{i}}az]{Isaule2021a}
Isaule, F.; Morera, I.; Massignan, P.; Juli{\'{a}}-D{\'{i}}az, B.
\newblock {Renormalization-group study of Bose polarons}.
\newblock {\em Phys. Rev. A} {\bf 2021}, {\em 104},~1--16,
  \href{http://xxx.lanl.gov/abs/2105.10801}{{\normalfont [2105.10801]}}.
\newblock
  doi:{\changeurlcolor{black}\href{https://doi.org/10.1103/PhysRevA.104.023317}{\detokenize{10.1103/PhysRevA.104.023317}}}.

\bibitem[Brauneis \em{et~al.}(2021)Brauneis, Hammer, Lemeshko, and
  Volosniev]{Brauneis2021}
Brauneis, F.; Hammer, H.W.; Lemeshko, M.; Volosniev, A.G.
\newblock {Impurities in a one-dimensional Bose gas: The flow equation
  approach}.
\newblock {\em SciPost Phys.} {\bf 2021}, {\em 11},~8,
  \href{http://xxx.lanl.gov/abs/2101.10958}{{\normalfont [2101.10958]}}.
\newblock
  doi:{\changeurlcolor{black}\href{https://doi.org/10.21468/SCIPOSTPHYS.11.1.008}{\detokenize{10.21468/SCIPOSTPHYS.11.1.008}}}.

\bibitem[Mistakidis \em{et~al.}(2019)Mistakidis, Volosniev, Zinner, and
  Schmelcher]{Mistakidis2019}
Mistakidis, S.I.; Volosniev, A.G.; Zinner, N.T.; Schmelcher, P.
\newblock {Effective approach to impurity dynamics in one-dimensional trapped
  Bose gases}.
\newblock {\em Phys. Rev. A} {\bf 2019}, {\em 100},
  \href{http://xxx.lanl.gov/abs/1809.01889}{{\normalfont [1809.01889]}}.
\newblock
  doi:{\changeurlcolor{black}\href{https://doi.org/10.1103/PhysRevA.100.013619}{\detokenize{10.1103/PhysRevA.100.013619}}}.

\bibitem[Grusdt \em{et~al.}(2017)Grusdt, Astrakharchik, and Demler]{Grusdt2017}
Grusdt, F.; Astrakharchik, G.E.; Demler, E.
\newblock {Bose polarons in ultracold atoms in one dimension: Beyond the
  Fr{\"{o}}hlich paradigm}.
\newblock {\em New J. Phys.} {\bf 2017}, {\em 19},~103035,
  \href{http://xxx.lanl.gov/abs/1704.02606}{{\normalfont [1704.02606]}}.
\newblock
  doi:{\changeurlcolor{black}\href{https://doi.org/10.1088/1367-2630/aa8a2e}{\detokenize{10.1088/1367-2630/aa8a2e}}}.

\bibitem[Ardila and Giorgini(2015)]{Ardila2015}
Ardila, L.A.; Giorgini, S.
\newblock {Impurity in a Bose-Einstein condensate: Study of the attractive and
  repulsive branch using quantum Monte Carlo methods}.
\newblock {\em Phys. Rev. A} {\bf 2015}, {\em 92},~1--12,
  \href{http://xxx.lanl.gov/abs/1507.07427}{{\normalfont [1507.07427]}}.
\newblock
  doi:{\changeurlcolor{black}\href{https://doi.org/10.1103/PhysRevA.92.033612}{\detokenize{10.1103/PhysRevA.92.033612}}}.

\bibitem[{Pe{\~{n}}a Ardila} \em{et~al.}(2019){Pe{\~{n}}a Ardila},
  J{\o}rgensen, Pohl, Giorgini, Bruun, and Arlt]{PenaArdila2019a}
{Pe{\~{n}}a Ardila}, L.A.; J{\o}rgensen, N.B.; Pohl, T.; Giorgini, S.; Bruun,
  G.M.; Arlt, J.J.
\newblock {Analyzing a Bose polaron across resonant interactions}.
\newblock {\em Phys. Rev. A} {\bf 2019}, {\em 99},~1--8,
  \href{http://xxx.lanl.gov/abs/1812.04609}{{\normalfont [1812.04609]}}.
\newblock
  doi:{\changeurlcolor{black}\href{https://doi.org/10.1103/PhysRevA.99.063607}{\detokenize{10.1103/PhysRevA.99.063607}}}.

\bibitem[Parisi and Giorgini(2017)]{Parisi2017}
Parisi, L.; Giorgini, S.
\newblock {Quantum Monte Carlo study of the Bose-polaron problem in a
  one-dimensional gas with contact interactions}.
\newblock {\em Phys. Rev. A} {\bf 2017}, {\em 95},~23619,
  \href{http://xxx.lanl.gov/abs/1612.01322}{{\normalfont [1612.01322]}}.
\newblock
  doi:{\changeurlcolor{black}\href{https://doi.org/10.1103/PhysRevA.95.023619}{\detokenize{10.1103/PhysRevA.95.023619}}}.

\bibitem[Schmidt and Enss(2021)]{Schmidt2021a}
Schmidt, R.; Enss, T.
\newblock {Self-stabilized Bose polarons} {\bf 2021}.
\newblock  \href{http://xxx.lanl.gov/abs/2102.13616}{{\normalfont
  [2102.13616]}}.

\bibitem[Ristivojevic(2021)]{Ristivojevic2021a}
Ristivojevic, Z.
\newblock {Dispersion relation of a polaron in the Yang-Gaudin Bose gas} {\bf
  2021}.
\newblock  \href{http://xxx.lanl.gov/abs/2111.10421}{{\normalfont
  [2111.10421]}}.

\bibitem[Lamacraft(2009)]{Lamacraft2009}
Lamacraft, A.
\newblock {Dispersion relation and spectral function of an impurity in a
  one-dimensional quantum liquid}.
\newblock {\em Phys. Rev. B} {\bf 2009}, {\em 79}.
\newblock
  doi:{\changeurlcolor{black}\href{https://doi.org/10.1103/PhysRevB.79.241105}{\detokenize{10.1103/PhysRevB.79.241105}}}.

\bibitem[Kulish \em{et~al.}(1976)Kulish, Manakov, and Faddeev]{Kulish1976}
Kulish, P.P.; Manakov, S.V.; Faddeev, L.D.
\newblock {Comparison of the exact quantum and quasiclassical results for a
  nonlinear Schr{\"{o}}dinger equation}.
\newblock {\em Theor. Math. Phys.} {\bf 1976}, {\em 28},~615--620.
\newblock
  doi:{\changeurlcolor{black}\href{https://doi.org/10.1007/BF01028912}{\detokenize{10.1007/BF01028912}}}.

\bibitem[Kanamoto \em{et~al.}(2008)Kanamoto, Carr, and Ueda]{Kanamoto2008}
Kanamoto, R.; Carr, L.D.; Ueda, M.
\newblock {Topological winding and unwinding in metastable Bose-Einstein
  condensates}.
\newblock {\em Phys. Rev. Lett.} {\bf 2008}, {\em 100},~060401.
\newblock
  doi:{\changeurlcolor{black}\href{https://doi.org/10.1103/PhysRevLett.100.060401}{\detokenize{10.1103/PhysRevLett.100.060401}}}.

\bibitem[Kanamoto \em{et~al.}(2010)Kanamoto, Carr, and Ueda]{Kanamoto2010}
Kanamoto, R.; Carr, L.D.; Ueda, M.
\newblock {Metastable quantum phase transitions in a periodic one-dimensional
  Bose gas. II. Many-body theory}.
\newblock {\em Phys. Rev. A} {\bf 2010}, {\em 81},~023625,
  \href{http://xxx.lanl.gov/abs/0910.2805}{{\normalfont [0910.2805]}}.
\newblock
  doi:{\changeurlcolor{black}\href{https://doi.org/10.1103/PhysRevA.81.023625}{\detokenize{10.1103/PhysRevA.81.023625}}}.

\bibitem[Jackson \em{et~al.}(2011)Jackson, Smyrnakis, Magiropoulos, and
  Kavoulakis]{Jackson2011}
Jackson, A.D.; Smyrnakis, J.; Magiropoulos, M.; Kavoulakis, G.M.
\newblock {Solitary waves and yrast states in Bose-Einstein condensed gases of
  atoms}.
\newblock {\em EPL} {\bf 2011}, {\em 95},~0--5,
  \href{http://xxx.lanl.gov/abs/1012.1816}{{\normalfont [1012.1816]}}.
\newblock
  doi:{\changeurlcolor{black}\href{https://doi.org/10.1209/0295-5075/95/30002}{\detokenize{10.1209/0295-5075/95/30002}}}.

\bibitem[Fialko \em{et~al.}(2012)Fialko, Delattre, Brand, and
  Kolovsky]{Fialko2012}
Fialko, O.; Delattre, M.C.; Brand, J.; Kolovsky, A.R.
\newblock {Nucleation in finite topological systems during continuous
  metastable quantum phase transitions}.
\newblock {\em Phys. Rev. Lett.} {\bf 2012}, {\em 108},~250402.
\newblock
  doi:{\changeurlcolor{black}\href{https://doi.org/10.1103/PhysRevLett.108.250402}{\detokenize{10.1103/PhysRevLett.108.250402}}}.

\bibitem[Sato \em{et~al.}(2012)Sato, Kanamoto, Kaminishi, and
  Deguchi]{Sato2012}
Sato, J.; Kanamoto, R.; Kaminishi, E.; Deguchi, T.
\newblock {Exact relaxation dynamics of a localized many-body state in the 1D
  bose gas}.
\newblock {\em Phys. Rev. Lett.} {\bf 2012}, {\em 108},~110401,
  \href{http://xxx.lanl.gov/abs/1112.4244}{{\normalfont [1112.4244]}}.
\newblock
  doi:{\changeurlcolor{black}\href{https://doi.org/10.1103/PhysRevLett.108.110401}{\detokenize{10.1103/PhysRevLett.108.110401}}}.

\bibitem[Syrwid and Sacha(2015)]{Syrwid2015}
Syrwid, A.; Sacha, K.
\newblock {Lieb-Liniger model: Emergence of dark solitons in the course of
  measurements of particle positions}.
\newblock {\em Phys. Rev. A} {\bf 2015}, {\em 92},~032110,
  \href{http://xxx.lanl.gov/abs/1505.06586}{{\normalfont [1505.06586]}}.
\newblock
  doi:{\changeurlcolor{black}\href{https://doi.org/10.1103/PhysRevA.92.032110}{\detokenize{10.1103/PhysRevA.92.032110}}}.

\bibitem[Shamailov and Brand(2019)]{Shamailov2019}
Shamailov, S.S.; Brand, J.
\newblock {Quantum dark solitons in the one-dimensional Bose gas}.
\newblock {\em Phys. Rev. A} {\bf 2019}, {\em 99},~43632,
  \href{http://xxx.lanl.gov/abs/1805.07856}{{\normalfont [1805.07856]}}.
\newblock
  doi:{\changeurlcolor{black}\href{https://doi.org/10.1103/PhysRevA.99.043632}{\detokenize{10.1103/PhysRevA.99.043632}}}.

\bibitem[Tsuzuki(1971)]{Tsuzuki1971}
Tsuzuki, T.
\newblock {Nonlinear waves in the Pitaevskii-Gross equation}.
\newblock {\em J. Low Temp. Phys.} {\bf 1971}, {\em 4},~441--457.
\newblock
  doi:{\changeurlcolor{black}\href{https://doi.org/10.1007/BF00628744}{\detokenize{10.1007/BF00628744}}}.

\bibitem[Shamailov and Brand(2016)]{Shamailov2016}
Shamailov, S.S.; Brand, J.
\newblock {Dark-soliton-like excitations in the Yang-Gaudin gas of attractively
  interacting fermions}.
\newblock {\em New J. Phys.} {\bf 2016}, {\em 18},~075004,
  \href{http://xxx.lanl.gov/abs/1603.04864}{{\normalfont [1603.04864]}}.
\newblock
  doi:{\changeurlcolor{black}\href{https://doi.org/10.1088/1367-2630/18/7/075004}{\detokenize{10.1088/1367-2630/18/7/075004}}}.

\bibitem[Syrwid(2021)]{Syrwid2021a}
Syrwid, A.
\newblock {Quantum dark solitons in ultracold one-dimensional Bose and Fermi
  gases},  2021,  \href{http://xxx.lanl.gov/abs/2009.12554}{{\normalfont
  [2009.12554]}}.
\newblock
  doi:{\changeurlcolor{black}\href{https://doi.org/10.1088/1361-6455/abd37f}{\detokenize{10.1088/1361-6455/abd37f}}}.

\bibitem[Astrakharchik and Brouzos(2013)]{Astrakharchik2013}
Astrakharchik, G.E.; Brouzos, I.
\newblock {Trapped one-dimensional ideal Fermi gas with a single impurity}.
\newblock {\em Phys. Rev. A} {\bf 2013}, {\em 88},~21602,
  \href{http://xxx.lanl.gov/abs/1303.7007}{{\normalfont [1303.7007]}}.
\newblock
  doi:{\changeurlcolor{black}\href{https://doi.org/10.1103/PhysRevA.88.021602}{\detokenize{10.1103/PhysRevA.88.021602}}}.

\bibitem[Booth \em{et~al.}(2009)Booth, Thom, and Alavi]{Booth2009}
Booth, G.H.; Thom, A.J.; Alavi, A.
\newblock {Fermion monte carlo without fixed nodes: A game of life, death, and
  annihilation in Slater determinant space}.
\newblock {\em J. Chem. Phys.} {\bf 2009}, {\em 131},~054106.
\newblock
  doi:{\changeurlcolor{black}\href{https://doi.org/10.1063/1.3193710}{\detokenize{10.1063/1.3193710}}}.

\bibitem[Cleland \em{et~al.}(2010)Cleland, Booth, and Alavi]{Cleland2010}
Cleland, D.; Booth, G.H.; Alavi, A.
\newblock {Communications: Survival of the fittest: Accelerating convergence in
  full configuration-interaction quantum Monte Carlo}.
\newblock {\em J. Chem. Phys.} {\bf 2010}, {\em 132},~41103.
\newblock
  doi:{\changeurlcolor{black}\href{https://doi.org/10.1063/1.3302277}{\detokenize{10.1063/1.3302277}}}.

\bibitem[Kalos and Whitlock(2009)]{Kalos2007}
Kalos, M.H.; Whitlock, P.A.
\newblock {\em {Monte Carlo Methods: Second Edition}}; Wiley-VCH Verlag GmbH \&
  Co. KGaA: Weinheim, Germany,  2009; pp. 1--203.
\newblock
  doi:{\changeurlcolor{black}\href{https://doi.org/10.1002/9783527626212}{\detokenize{10.1002/9783527626212}}}.

\bibitem[Spencer \em{et~al.}(2012)Spencer, Blunt, and Foulkes]{Spencer2012}
Spencer, J.S.; Blunt, N.S.; Foulkes, W.M.
\newblock {The sign problem and population dynamics in the full configuration
  interaction quantum Monte Carlo method}.
\newblock {\em J. Chem. Phys.} {\bf 2012}, {\em 136},~054110,
  \href{http://xxx.lanl.gov/abs/1110.5479}{{\normalfont [1110.5479]}}.
\newblock
  doi:{\changeurlcolor{black}\href{https://doi.org/10.1063/1.3681396}{\detokenize{10.1063/1.3681396}}}.

\bibitem[Booth \em{et~al.}(2011)Booth, Cleland, Thom, and Alavi]{Booth2011}
Booth, G.H.; Cleland, D.; Thom, A.J.; Alavi, A.
\newblock {Breaking the carbon dimer: The challenges of multiple bond
  dissociation with full configuration interaction quantum Monte Carlo
  methods}.
\newblock {\em J. Chem. Phys.} {\bf 2011}, {\em 135},~84104.
\newblock
  doi:{\changeurlcolor{black}\href{https://doi.org/10.1063/1.3624383}{\detokenize{10.1063/1.3624383}}}.

\bibitem[Cleland \em{et~al.}(2012)Cleland, Booth, Overy, and
  Alavi]{Cleland2012a}
Cleland, D.; Booth, G.H.; Overy, C.; Alavi, A.
\newblock {Taming the first-row diatomics: A full configuration interaction
  quantum Monte Carlo study}.
\newblock {\em J. Chem. Theory Comput.} {\bf 2012}, {\em 8},~4138--4152.
\newblock
  doi:{\changeurlcolor{black}\href{https://doi.org/10.1021/ct300504f}{\detokenize{10.1021/ct300504f}}}.

\bibitem[Booth \em{et~al.}(2013)Booth, Gr{\"{u}}neis, Kresse, and
  Alavi]{Booth2013}
Booth, G.H.; Gr{\"{u}}neis, A.; Kresse, G.; Alavi, A.
\newblock {Towards an exact description of electronic wavefunctions in real
  solids}.
\newblock {\em Nature} {\bf 2013}, {\em 493},~365--370.
\newblock
  doi:{\changeurlcolor{black}\href{https://doi.org/10.1038/nature11770}{\detokenize{10.1038/nature11770}}}.

\bibitem[Schwarz \em{et~al.}(2015)Schwarz, Booth, and Alavi]{Schwarz2015}
Schwarz, L.R.; Booth, G.H.; Alavi, A.
\newblock {Insights into the structure of many-electron wave functions of
  Mott-insulating antiferromagnets: The three-band Hubbard model in full
  configuration interaction quantum Monte Carlo}.
\newblock {\em Phys. Rev. B} {\bf 2015}, {\em 91},~45139.
\newblock
  doi:{\changeurlcolor{black}\href{https://doi.org/10.1103/PhysRevB.91.045139}{\detokenize{10.1103/PhysRevB.91.045139}}}.

\bibitem[Yun \em{et~al.}(2017)Yun, Dong, and Zhu]{Yun2017}
Yun, S.J.; Dong, T.K.; Zhu, S.N.
\newblock {Validation of the Ability of Full Configuration Interaction Quantum
  Monte Carlo for Studying the 2D Hubbard Model}.
\newblock {\em Chinese Phys. Lett.} {\bf 2017}, {\em 34},~0--5.
\newblock
  doi:{\changeurlcolor{black}\href{https://doi.org/10.1088/0256-307X/34/8/080201}{\detokenize{10.1088/0256-307X/34/8/080201}}}.

\bibitem[Yun \em{et~al.}(2021)Yun, Dobrautz, Luo, and Alavi]{Yun2021}
Yun, S.; Dobrautz, W.; Luo, H.; Alavi, A.
\newblock {Benchmark study of Nagaoka ferromagnetism by spin-adapted full
  configuration interaction quantum Monte Carlo}.
\newblock {\em Phys. Rev. B} {\bf 2021}, {\em 104},
  \href{http://xxx.lanl.gov/abs/2105.06802}{{\normalfont [2105.06802]}}.
\newblock
  doi:{\changeurlcolor{black}\href{https://doi.org/10.1103/physrevb.104.235102}{\detokenize{10.1103/physrevb.104.235102}}}.

\bibitem[Ebling \em{et~al.}(2021)Ebling, Alavi, and Brand]{Ebling2021a}
Ebling, U.; Alavi, A.; Brand, J.
\newblock {Signatures of the BCS-BEC crossover in the yrast spectra of Fermi
  quantum rings}.
\newblock {\em Phys. Rev. Res.} {\bf 2021}, {\em 3},~23142,
  \href{http://xxx.lanl.gov/abs/2011.14538}{{\normalfont [2011.14538]}}.
\newblock
  doi:{\changeurlcolor{black}\href{https://doi.org/10.1103/PhysRevResearch.3.023142}{\detokenize{10.1103/PhysRevResearch.3.023142}}}.

\bibitem[Yang \em{et~al.}(2020)Yang, Pahl, and Brand]{Yang2020}
Yang, M.; Pahl, E.; Brand, J.
\newblock {Improved walker population control for full configuration
  interaction quantum Monte Carlo}.
\newblock {\em J. Chem. Phys.} {\bf 2020}, {\em 153},~174103,
  \href{http://xxx.lanl.gov/abs/2008.01927}{{\normalfont [2008.01927]}}.
\newblock
  doi:{\changeurlcolor{black}\href{https://doi.org/10.1063/5.0023088}{\detokenize{10.1063/5.0023088}}}.

\bibitem[Brand \em{et~al.}(2021)Brand, Yang, and Pahl]{Brand2021}
Brand, J.; Yang, M.; Pahl, E.
\newblock {Stochastic differential equation approach to understanding the
  population control bias in full configuration interaction quantum Monte
  Carlo} {\bf 2021}.
\newblock  \href{http://xxx.lanl.gov/abs/2103.07800}{{\normalfont
  [2103.07800]}}.

\bibitem[Castin(2004)]{Castin2004}
Castin, Y.
\newblock {Simple theoretical tools for low dimension Bose gases}.
\newblock {\em J. Phys. IV JP} {\bf 2004}, {\em 116},~89--132,
  \href{http://xxx.lanl.gov/abs/0407118}{{\normalfont
  [arXiv:cond-mat/0407118]}}.
\newblock
  doi:{\changeurlcolor{black}\href{https://doi.org/10.1051/jp4:2004116004}{\detokenize{10.1051/jp4:2004116004}}}.

\bibitem[Ernst \em{et~al.}(2011)Ernst, Hallwood, Gulliksen, Meyer, and
  Brand]{Ernst2011}
Ernst, T.; Hallwood, D.W.; Gulliksen, J.; Meyer, H.D.; Brand, J.
\newblock {Simulating strongly correlated multiparticle systems in a truncated
  Hilbert space}.
\newblock {\em Phys. Rev. A} {\bf 2011}, {\em 84},~23623.
\newblock
  doi:{\changeurlcolor{black}\href{https://doi.org/10.1103/PhysRevA.84.023623}{\detokenize{10.1103/PhysRevA.84.023623}}}.

\bibitem[Vigor \em{et~al.}(2015)Vigor, Spencer, Bearpark, and Thom]{Vigor2015}
Vigor, W.A.; Spencer, J.S.; Bearpark, M.J.; Thom, A.J.
\newblock {Minimising biases in full configuration interaction quantum Monte
  Carlo}.
\newblock {\em J. Chem. Phys.} {\bf 2015}, {\em 142},~104101,
  \href{http://xxx.lanl.gov/abs/1407.1753}{{\normalfont [1407.1753]}}.
\newblock
  doi:{\changeurlcolor{black}\href{https://doi.org/10.1063/1.4913644}{\detokenize{10.1063/1.4913644}}}.

\bibitem[Lim and Weare(2017)]{Lim2015}
Lim, L.H.; Weare, J.
\newblock {Fast randomized iteration: Diffusion Monte Carlo through the lens of
  numerical linear algebra}.
\newblock {\em SIAM Rev.} {\bf 2017}, {\em 59},~547--587,
  \href{http://xxx.lanl.gov/abs/1508.06104}{{\normalfont [1508.06104]}}.
\newblock
  doi:{\changeurlcolor{black}\href{https://doi.org/10.1137/15M1040827}{\detokenize{10.1137/15M1040827}}}.

\bibitem[Greene \em{et~al.}(2019)Greene, Webber, Weare, and
  Berkelbach]{Greene2019}
Greene, S.M.; Webber, R.J.; Weare, J.; Berkelbach, T.C.
\newblock {Beyond Walkers in Stochastic Quantum Chemistry: Reducing Error Using
  Fast Randomized Iteration}.
\newblock {\em J. Chem. Theory Comput.} {\bf 2019}, {\em 15},~4834--4850,
  \href{http://xxx.lanl.gov/abs/1905.00995}{{\normalfont [1905.00995]}}.
\newblock
  doi:{\changeurlcolor{black}\href{https://doi.org/10.1021/acs.jctc.9b00422}{\detokenize{10.1021/acs.jctc.9b00422}}}.

\bibitem[Greene \em{et~al.}(2020)Greene, Webber, Weare, and
  Berkelbach]{Greene2020}
Greene, S.M.; Webber, R.J.; Weare, J.; Berkelbach, T.C.
\newblock {Improved Fast Randomized Iteration Approach to Full Configuration
  Interaction}.
\newblock {\em J. Chem. Theory Comput.} {\bf 2020}, {\em 16},~5572--5585,
  \href{http://xxx.lanl.gov/abs/2005.00654}{{\normalfont [2005.00654]}}.
\newblock
  doi:{\changeurlcolor{black}\href{https://doi.org/10.1021/acs.jctc.0c00437}{\detokenize{10.1021/acs.jctc.0c00437}}}.

\bibitem[{\v{C}ufar} \em{et~al.}(2022){\v{C}ufar}, Pahl, and
  Brand]{Cufar2022ep}
{\v{C}ufar}, M.; Pahl, E.; Brand, J.
\newblock Efficient sampling algorithms for FCIQMC.
\newblock to be published.

\bibitem[Bezanson \em{et~al.}(2017)Bezanson, Edelman, Karpinski, and
  Shah]{Bezanson2017}
Bezanson, J.; Edelman, A.; Karpinski, S.; Shah, V.B.
\newblock {Julia: A fresh approach to numerical computing}.
\newblock {\em SIAM Rev.} {\bf 2017}, {\em 59},~65--98,
  \href{http://xxx.lanl.gov/abs/1411.1607}{{\normalfont [1411.1607]}}.
\newblock
  doi:{\changeurlcolor{black}\href{https://doi.org/10.1137/141000671}{\detokenize{10.1137/141000671}}}.

\bibitem[rim()]{rimucode}
\texttt{Rimu.jl} is available at \url{https://github.com/joachimbrand/Rimu.jl}.

\bibitem[Booth \em{et~al.}(2014)Booth, Smart, and Alavi]{Booth2014}
Booth, G.H.; Smart, S.D.; Alavi, A.
\newblock {Linear-scaling and parallelisable algorithms for stochastic quantum
  chemistry}.
\newblock {\em Mol. Phys.} {\bf 2014}, {\em 112},~1855--1869,
  \href{http://xxx.lanl.gov/abs/1305.6981}{{\normalfont [1305.6981]}}.
\newblock
  doi:{\changeurlcolor{black}\href{https://doi.org/10.1080/00268976.2013.877165}{\detokenize{10.1080/00268976.2013.877165}}}.

\bibitem[Clement and Quinn(1993)]{mpi1993}
Clement, M.J.; Quinn, M.J.
\newblock Analytical Performance Prediction on Multicomputers.
\newblock  Proceedings of the 1993 ACM/IEEE Conference on Supercomputing;
  Association for Computing Machinery: New York, NY, USA,  1993; Supercomputing
  '93, p. 886–894.
\newblock
  doi:{\changeurlcolor{black}\href{https://doi.org/10.1145/169627.169856}{\detokenize{10.1145/169627.169856}}}.

\bibitem[Byrne \em{et~al.}(2021)Byrne, Wilcox, and Churavy]{Byrne2021}
Byrne, S.; Wilcox, L.C.; Churavy, V.
\newblock {MPI.jl: Julia bindings for the Message Passing Interface}.
\newblock {\em JuliaCon Proc.} {\bf 2021}, {\em 1},~68.
\newblock
  doi:{\changeurlcolor{black}\href{https://doi.org/10.21105/jcon.00068}{\detokenize{10.21105/jcon.00068}}}.

\bibitem[Flyvbjerg and Petersen(1989)]{Flyvbjerg1989}
Flyvbjerg, H.; Petersen, H.G.
\newblock {Error estimates on averages of correlated data}.
\newblock {\em J. Chem. Phys.} {\bf 1989}, {\em 91},~461--466.
\newblock
  doi:{\changeurlcolor{black}\href{https://doi.org/10.1063/1.457480}{\detokenize{10.1063/1.457480}}}.

\bibitem[Jonsson(2018)]{Jonsson2018}
Jonsson, M.
\newblock {Standard error estimation by an automated blocking method}.
\newblock {\em Phys. Rev. E} {\bf 2018}, {\em 98},~043304.
\newblock
  doi:{\changeurlcolor{black}\href{https://doi.org/10.1103/PhysRevE.98.043304}{\detokenize{10.1103/PhysRevE.98.043304}}}.

\bibitem[Overy \em{et~al.}(2014)Overy, Booth, Blunt, Shepherd, Cleland, and
  Alavi]{Overy2014}
Overy, C.; Booth, G.H.; Blunt, N.S.; Shepherd, J.J.; Cleland, D.; Alavi, A.
\newblock {Unbiased reduced density matrices and electronic properties from
  full configuration interaction quantum Monte Carlo}.
\newblock {\em J. Chem. Phys.} {\bf 2014}, {\em 141},~244117.
\newblock
  doi:{\changeurlcolor{black}\href{https://doi.org/10.1063/l.4904313}{\detokenize{10.1063/l.4904313}}}.

\bibitem[Lieb(1963)]{lieb63:2}
Lieb, E.H.
\newblock {Exact analysis of an interacting bose gas. II. the excitation
  spectrum}.
\newblock {\em Phys. Rev.} {\bf 1963}, {\em 130},~1616--1624.
\newblock
  doi:{\changeurlcolor{black}\href{https://doi.org/10.1103/PhysRev.130.1616}{\detokenize{10.1103/PhysRev.130.1616}}}.

\bibitem[Konotop and Pitaevskii(2004)]{Konotop2004}
Konotop, V.V.; Pitaevskii, L.
\newblock {Landau dynamics of a grey soliton in a trapped condensate}.
\newblock {\em Phys. Rev. Lett.} {\bf 2004}, {\em 93},~8--11,
  \href{http://xxx.lanl.gov/abs/0408660}{{\normalfont
  [arXiv:cond-mat/0408660]}}.
\newblock
  doi:{\changeurlcolor{black}\href{https://doi.org/10.1103/PhysRevLett.93.240403}{\detokenize{10.1103/PhysRevLett.93.240403}}}.

\bibitem[Astrakharchik and Pitaevskii(2012)]{Astrakharchik2012}
Astrakharchik, G.E.; Pitaevskii, L.P.
\newblock {Lieb's soliton-like excitations in harmonic trap}.
\newblock {\em EPL} {\bf 2012}, {\em 102},~30004,
  \href{http://xxx.lanl.gov/abs/1210.8337}{{\normalfont [1210.8337]}}.
\newblock
  doi:{\changeurlcolor{black}\href{https://doi.org/10.1209/0295-5075/102/30004}{\detokenize{10.1209/0295-5075/102/30004}}}.

\bibitem[Anderson(1966)]{Anderson1966}
Anderson, P.W.
\newblock {Considerations on the Flow of Superfluid Helium}.
\newblock {\em Rev. Mod. Phys.} {\bf 1966}, {\em 38},~298--310.
\newblock
  doi:{\changeurlcolor{black}\href{https://doi.org/10.1103/RevModPhys.38.298}{\detokenize{10.1103/RevModPhys.38.298}}}.

\bibitem[Jeszenszki \em{et~al.}(2018)Jeszenszki, Luo, Alavi, and
  Brand]{Jeszenszki2018}
Jeszenszki, P.; Luo, H.; Alavi, A.; Brand, J.
\newblock {Accelerating the convergence of exact diagonalization with the
  transcorrelated method: Quantum gas in one dimension with contact
  interactions}.
\newblock {\em Phys. Rev. A} {\bf 2018}, {\em 98},~53627,
  \href{http://xxx.lanl.gov/abs/1806.11268}{{\normalfont [1806.11268]}}.
\newblock
  doi:{\changeurlcolor{black}\href{https://doi.org/10.1103/PhysRevA.98.053627}{\detokenize{10.1103/PhysRevA.98.053627}}}.

\bibitem[Jeszenszki \em{et~al.}(2020)Jeszenszki, Ebling, Luo, Alavi, and
  Brand]{Jeszenszki2020}
Jeszenszki, P.; Ebling, U.; Luo, H.; Alavi, A.; Brand, J.
\newblock {Eliminating the wave-function singularity for ultracold atoms by a
  similarity transformation}.
\newblock {\em Phys. Rev. Res.} {\bf 2020}, {\em 2},~43270,
  \href{http://xxx.lanl.gov/abs/2002.05987}{{\normalfont [2002.05987]}}.
\newblock
  doi:{\changeurlcolor{black}\href{https://doi.org/10.1103/PhysRevResearch.2.043270}{\detokenize{10.1103/PhysRevResearch.2.043270}}}.

\bibitem[Keiler \em{et~al.}(2021)Keiler, Mistakidis, and
  Schmelcher]{Keiler2020}
Keiler, K.; Mistakidis, S.I.; Schmelcher, P.
\newblock {Polarons and their induced interactions in highly imbalanced triple
  mixtures}.
\newblock {\em Phys. Rev. A} {\bf 2021}, {\em 104},
  \href{http://xxx.lanl.gov/abs/2012.04034}{{\normalfont [2012.04034]}}.
\newblock
  doi:{\changeurlcolor{black}\href{https://doi.org/10.1103/PhysRevA.104.L031301}{\detokenize{10.1103/PhysRevA.104.L031301}}}.

\bibitem[Camacho-Guardian \em{et~al.}(2018)Camacho-Guardian, {Pe{\~{n}}a
  Ardila}, Pohl, and Bruun]{Camacho-Guardian2018}
Camacho-Guardian, A.; {Pe{\~{n}}a Ardila}, L.A.; Pohl, T.; Bruun, G.M.
\newblock {Bipolarons in a Bose-Einstein Condensate}.
\newblock {\em Phys. Rev. Lett.} {\bf 2018}, {\em 121},~13401,
  \href{http://xxx.lanl.gov/abs/1804.00402}{{\normalfont [1804.00402]}}.
\newblock
  doi:{\changeurlcolor{black}\href{https://doi.org/10.1103/PhysRevLett.121.013401}{\detokenize{10.1103/PhysRevLett.121.013401}}}.

\bibitem[Will \em{et~al.}(2021)Will, Astrakharchik, and Fleischhauer]{Will2021}
Will, M.; Astrakharchik, G.E.; Fleischhauer, M.
\newblock {Polaron Interactions and Bipolarons in One-Dimensional Bose Gases in
  the Strong Coupling Regime}.
\newblock {\em Phys. Rev. Lett.} {\bf 2021}, {\em 127},
  \href{http://xxx.lanl.gov/abs/2101.11997}{{\normalfont [2101.11997]}}.
\newblock
  doi:{\changeurlcolor{black}\href{https://doi.org/10.1103/PhysRevLett.127.103401}{\detokenize{10.1103/PhysRevLett.127.103401}}}.

\bibitem[Petkovic and Ristivojevic(2021)]{Petkovic2021}
Petkovic, A.; Ristivojevic, Z.
\newblock {Mediated interaction between polarons in a one-dimensional Bose gas}
  {\bf 2021}.
\newblock  \href{http://xxx.lanl.gov/abs/2103.08772}{{\normalfont
  [2103.08772]}}.

\bibitem[Umrigar \em{et~al.}(1993)Umrigar, Nightingale, and Runge]{Umrigar1993}
Umrigar, C.J.; Nightingale, M.P.; Runge, K.J.
\newblock {A diffusion Monte Carlo algorithm with very small time-step errors}.
\newblock {\em J. Chem. Phys.} {\bf 1993}, {\em 99},~2865--2890.
\newblock
  doi:{\changeurlcolor{black}\href{https://doi.org/10.1063/1.465195}{\detokenize{10.1063/1.465195}}}.

\bibitem[Ghanem \em{et~al.}(2021)Ghanem, Liebermann, and Alavi]{Ghanem2021}
Ghanem, K.; Liebermann, N.; Alavi, A.
\newblock {Population control bias and importance sampling in full
  configuration interaction quantum Monte Carlo}.
\newblock {\em Phys. Rev. B} {\bf 2021}, {\em 103},
  \href{http://xxx.lanl.gov/abs/2102.11016}{{\normalfont [2102.11016]}}.
\newblock
  doi:{\changeurlcolor{black}\href{https://doi.org/10.1103/PhysRevB.103.155135}{\detokenize{10.1103/PhysRevB.103.155135}}}.

\end{thebibliography}

\end{document}